\documentclass[aps,prd,preprint,groupaddress,amssymb,amsmath,floatfix,preprintnumbers,nofootinbib,altaffilletter]{revtex4-1}
\pdfoutput=1
\usepackage[utf8]{inputenc}
\usepackage[OT1]{fontenc}
\usepackage{yfonts,amsmath,amsthm,amsfonts,amssymb,amscd}
\usepackage{enumerate}
\usepackage{fancyhdr}
\usepackage{mathrsfs}
\usepackage{cancel}
\usepackage{slashed}
\usepackage[flushleft]{threeparttable}
\usepackage{makecell,booktabs}
\usepackage{braket}
\usepackage{relsize}
\usepackage{tikz}
\usepackage{pgfplots}
\usetikzlibrary{pgfplots.polar}
\usepackage[colorlinks=true, citecolor=blue, linkcolor=blue,urlcolor=blue]{hyperref}
\usepackage[margin=3cm]{geometry}
\usepackage[T1]{fontenc} % if needed
\begin{document} 
\count\footins = 1000

\title{\boldmath Muon $g-2$  in two Higgs doublet models with vectorlike leptons}

%% %simple case: 2 authors, same institution
%% \author{A. Uthor}
%% \author{and A. Nother Author}
%% \affiliation{Institution,\\Address, Country}

% more complex case: 4 authors, 3 institutions, 2 footnotes
\author{Radovan Dermisek}
\email{dermisek@indiana.edu}
\affiliation{Physics Department, Indiana University,\\Bloomington, IN, 47405, USA}
\author{Keith Hermanek}
\email{khermane@iu.edu}
\affiliation{Physics Department, Indiana University,\\Bloomington, IN, 47405, USA}
\author{Navin McGinnis}
\email{nmcginnis@triumf.ca}
\affiliation{TRIUMF, 4004 Westbrook Mall, Vancouver, BC, Canada V6T 2A3}
\affiliation{High Energy Physics Division, Argonne National Laboratory,\\Argonne, IL, 60439}
%\author[a,2]{T. Hird\note{Also at Some University.}}
%\author[a,2]{and Fourth}

% The "\note" macro will give a warning: "Ignoring empty anchor..."
% you can safely ignore it.

%\affiliation[c]{A School for Advanced Studies,\\some-location, Country}

% e-mail addresses: one for each author, in the same order as the authors

%\emailAdd{fourth@one.univ}

\begin{abstract}We calculate contributions to the anomalous magnetic moment of the muon from heavy neutral and charged Higgs bosons and new leptons in two Higgs doublet models extended by vectorlike leptons. We present detailed predictions of two models with type-II couplings to standard model fermions, motivated by a $Z_2$ symmetry and supersymmetry. In addition, we compare the results with the standard model extended by vectorlike leptons. We find that the model motivated by a $Z_{2}$ symmetry can generate much larger contributions to the magnetic moment compared to the standard model, even by two orders of magnitude due to $\tan^{2}\beta$ enhancement, while satisfying current constraints. As a consequence, the standard model explanation of the anomaly requires much larger corrections to muon couplings making this model easier to probe at future precision machines. Additionally, we find that the model with couplings motivated by supersymmetry typically leads to much smaller contributions to the magnetic moment as a result of cancellations. We also identify interesting scenarios where contributions from the charged Higgs boson can fully explain the anomaly.
\end{abstract}

\maketitle
%\tableofcontents

\section{Introduction}
\label{sec:intro}

The Standard Model (SM) provides a spectacular description of nature, surviving stringent tests at both the current energy and precision frontiers. Indeed the absence of any direct signal for new particles at the LHC implies strong bounds for many kinds of new particles up to several TeV. Further, the discovery of the Higgs boson and the subsequent measurements of the Higgs couplings to gauge bosons and fermions indicate that the SM is the appropriate effective theory of electroweak (EW) symmetry breaking. 

Despite the lack of any direct clue for new particles, some discrepancies with SM predictions still persist evoking a variety of models for new physics whose low-energy effects could be probed indirectly. In particular, the measurement of the magnetic moment of the muon deviates from the SM prediction by more than four standard deviations~\cite{Abi:2021gix, Aoyama:2020ynm,Zyla:2020zbs}. Examples of models which may lead to an explanation of this discrepancy with particles at or slightly above EW include possible new fermions, scalars, gauge bosons, or combinations of new particles, e.g. in the MSSM. For detailed reviews see~\cite{Czarnecki:2001pv, Endo:2013bba, Freitas:2014pua, Lindner:2016bgg,Endo:2020mqz} and references therein. Naively, new particles which can account for the anomalous magnetic moment cannot be too far above the EW scale, since the typical contribution from new particles can be parameterized by $\Delta a_{\mu}\simeq g_{NP}^2m_{\mu}^2/16\pi^2 m_{NP}^2$, where $g_{NP}$ and $m_{NP}$ are the coupling and mass of new particles. In some cases, certain enhancements can allow for heavier particles. For instance in the MSSM, the contribution can be enhanced by $\tan\beta$~\cite{Moroi:1995yh}. Alternative explanations involve very light particles which, to avoid a variety of constraints, must be singlets under the SM~\cite{Chen:2015vqy, Marciano:2016yhf,Davoudiasl:2018fbb, Liu:2018xkx, Bauer:2019gfk, Liu:2020qgx,Jana:2020pxx}. 

In models with new fermions which have the same quantum numbers as SM leptons, the contributions to $(g-2)_{\mu}$ associated to new physics are proportional to the mixing parameter, $m_{\mu}^{LE}$, which simultaneously contributes to the muon mass. The contribution to $(g-2)_{\mu}$ can be estimated by $\Delta a_{\mu} \simeq m_{\mu} \lambda^3v/16\pi^2m_{NP}^2 \simeq m_{\mu} m_{\mu}^{LE}/16\pi^2v^2$~\cite{Kannike:2011ng, Dermisek:2013gta}. In this case chirality flipping operators lead to a \textit{chiral enhancement}, $\lambda v/m_{\mu}$, compared to the typical contribution. Chiral enhancement effects related to $(g-2)_{\mu}$ are additionally motivated by connections with recent B anomalies~\cite{Raby:2017igl,Crivellin:2018qmi,Barman:2018jhz,Arnan:2019uhr,Kawamura:2019rth}, the Cabibbo angle anomaly~\cite{Endo:2020tkb,Crivellin:2020ebi}, and dark matter~\cite{ Kowalska:2017iqv, Calibbi:2018rzv,Jana:2020joi}.

In this paper, we focus mainly on type-II 2HDM models with vectorlike leptons as an explanation for the anomalous measurement of $(g-2)_{\mu}$. We study in detail a type-II 2HDM motivated by a $Z_{2}$ symmetry (2HDM-II-$Z_{2}$), highlights of which were presented previously in~\cite{Dermisek:2020cod}. A striking feature of this scenario is found in a $\tan^{2}\beta$ enhancement in the contributions of heavy Higgses to $(g-2)_{\mu}$ compared to those of $W,Z,$ and $h$ \textit{in addition to} the chiral enhancement expected in models with VL. In contrast, the same couplings which generate a large correction to $(g-2)_{\mu}$ also lead to corrections of $W,Z,$ and $h$ couplings to the muon resulting from mixing that are $\tan^{2}\beta$ suppressed. This would allow for a contribution to $\Delta a_{\mu}$ even two orders of magnitude larger than the measured value while simultaneously satisfying low-energy observables, or an explanation of the measured value, $\Delta a_{\mu}^{exp}$, with tiny corrections to SM couplings, or even an explanation of $\Delta a_{\mu}^{exp}$ from new leptons with masses of tens of TeV. Interestingly, future precision measurements can fully explore scenarios with heavy new leptons indirectly~\cite{Dermisek:2020cod}. In addition, a muon collider would be perfectly suited to explore heavy lepton masses directly~\cite{Capdevilla:2020qel,Yin:2020afe,Capdevilla:2021rwo}.

We also discuss a version of the 2HDM motivated by supersymmetry (2HDM-II-S). In particular, we call attention to the fact that in either model, the couplings of the Higgs doublets to SM leptons are indistinguishable. However, when the models are extended with VL each symmetry dictates a different structure of Yukawa couplings leading to drastically different results. In this version of the model we find that the contributions to $\Delta a_{\mu}$ from vectorlike leptons and heavy Higgses with comparable masses tend to cancel with those of $W,Z,$ and $h$. Viable explanations of $\Delta a_{\mu}^{exp}$ can be achieved either by decoupled heavy Higgses or from the charged Higgs contribution if vectorlike neutral singlets are included.

Furthermore, we extend previous studies of the SM extended with vectorlike leptons~\cite{Kannike:2011ng,Dermisek:2013gta}. In particular, we include couplings to vectorlike neutral singlets (also considered previously in~\cite{Dermisek:2014cia}), and extend the range of possible couplings and masses that can explain $\Delta a_{\mu}^{exp}$. In addition, we impose updated experimental constraints emphasizing the impact of recent measurements of the SM Higgs coupling to the muon~\cite{Aad:2020xfq}. It has been noted that the correlation of the Higgs coupling to the muon with other observables can often give complementary information on models for new physics~\cite{Kannike:2011ng,Dermisek:2013gta,Crivellin:2020tsz,Dermisek:2020cod}. Interestingly, we find that this constraint limits the possible contribution to $\Delta a_{\mu}$ in the SM with vectorlike leptons close to the current central value, while in the 2HDM-II-$Z_{2}$ it allows for even two orders of magnitude larger contribution to $\Delta a_{\mu}$ than the measured value. However, the 1$\sigma$ range of $\Delta a_{\mu}^{exp}$ can be explained with a similar range of heavy lepton masses as in the 2HDM-II-$Z_{2}$. To illustrate the impact of future precision measurements, we study possible modifications of $W,~Z$, and $h$ couplings to the muon.

In our discussion we focus on scenarios where vectorlike leptons share analogous quantum numbers to SM leptons. This allows for straightforward extensions of the SM by complete vectorlike families in the context of simple unified models. The extension of the SM with vectorlike familes provides a possible explanation for the observed hierarchy of gauge couplings~\cite{Dermisek:2012as,Dermisek:2012ke}, while the MSSM with a complete vectorlike family can explain the structure of the seven largest couplings in the SM at the EW scale when all new particles are in the multi-TeV range~\cite{Dermisek:2017ihj,Dermisek:2018hxq,Dermisek:2018ujw}. Vectorlike quarks around the same scales can also lead to more natural EW symmetry breaking~\cite{Dermisek:2016tzw,Cohen:2020ohi}. For other examples of explanations of $\Delta a_{\mu}^{exp}$ with vectorlike leptons either with the same or different quantum numbers, see also Refs.~\cite{Arnan:2016cpy,  Kowalska:2017iqv, Megias:2017dzd, Raby:2017igl, Crivellin:2018qmi, Hiller:2019mou, Kawamura:2019rth, Hiller:2020fbu, Endo:2020tkb, Frank:2020smf, Chun:2020uzw, Kowalska:2020zve}.\footnote{In particular, similar 2HDM variants with VL have been explored recently in \cite{Frank:2020smf,Chun:2020uzw}. We find disagreement with the results in~\cite{Chun:2020uzw} in connection with $(g-2)_{e}$, where the neutral Higgs contributions are incomplete. Further, the authors claim that the charged Higgs contribution does not have any chiral enhancement, which we do not find to be correct. In~\cite{Frank:2020smf} the authors do not consider chirally-enhanced one-loop contributions and rather solely consider two-loop Barr-Zee contributions to $(g-2)_{\mu}$. However, we find that these contributions are negligible compared to chirally-enhanced one-loop contributions by several orders of magnitude.}  Related studies of $(g-2)_{\mu}$ in the MSSM with vectorlike leptons (not including one-loop contributions from heavy Higgses) were presented in~\cite{Endo:2011xq,Endo:2011mc,Choudhury:2017fuu}. For previous studies of supersymmetric models with vectorlike leptons, see also~\cite{Joglekar:2013zya,Kyae:2013hda}. Related discussions of collider searches for heavy new leptons can be found in~\cite{Kumar:2015tna,Dermisek:2015oja,Dermisek:2015hue,Dermisek:2016via,Bhattiprolu:2019vdu, Bissmann:2020lge} and similar studies with vectorlike quarks in~\cite{Dermisek:2019vkc,Dermisek:2019heo,Dermisek:2020gbr}.

This paper is organized as follows. In section~\ref{sec:model}, we describe the 2HDM-II-$Z_{2}$, 2HDM-II-S, and SM extended with vectorlike leptons which mix with the muon at tree level. In section~\ref{sec:loops}, we present formulae for contributions to $(g-2)_{\mu}$ in models with extended Higgs and lepton sectors that can be applied to any model. We discuss details of our analysis and a variety of constraints relevant to heavy leptons and Higgs bosons in section~\ref{sec:constraints}. We present detailed results and discussion for all three models in section~\ref{sec:results} and conclude in section~\ref{sec:conclusions}. In Appendix~\ref{sec:Z2_appendix}, we provide general formulas for couplings of the muon to $Z$, $W$, and Higgs bosons in the 2HDM-II-$Z_{2}$ and provide an explicit derivation of the Goldstone boson equivalence theorem for couplings of the $Z$ and $W$ boson. In addition we list useful approximate formulas which aid in understanding of the results. We provide details of the 2HDM-II-S in Appendix~\ref{app:SUSY_model}. We comment on the relative size of possible Barr-Zee contributions in Appendix~\ref{app:BZ}.

\section{Models}
\label{sec:model}
We consider a two Higgs doublet model extended with vector-like leptons (VL) in which both $SU(2)$ doublet, $L_{L,R}$, and singlet representations, $E_{L,R}$ and $N_{L,R}$, are included. We assume that the left-handed new doublet, $L_{L}$, transforms under the same representations as the left-handed SM leptons. Likewise, the right-handed charged singlet, $E_{R}$, has the same quantum numbers as the right-handed SM leptons. Further, we assume couplings of SM leptons to the Higgs doublets as in  type-{\MakeUppercase{\romannumeral 2}} models where $H_{d}$ couples exclusively to the down-sector leptons and $H_{u}$ to the up-sector. This can be achieved by assigning appropriate charges under a $Z_{2}$ symmetry. Alternatively, the supersymmetric extension of the SM automatically leads to couplings of SM fermions of type-II~\cite{Gunion:1989we}. However, when VL are included the $Z_{2}$ symmetry and supersymmetry enforce different structures of their Yukawa couplings to the Higgs doublets, and thus we distinguish the two models. We will also compare these models with the SM extended with VL. In all cases, the leading contributions of the model to $(g-2)_{\mu}$ originate from possible mixing of VL leptons to the 2nd generation SM leptons. Thus, for simplicity we will consider only Yukawa couplings leading to mixing of VL leptons to the muon and muon neutrino. 

\subsection{2HDM-II-$Z_{2}$ with vectorlike leptons}
For the main focus of this paper, we consider the type-II two Higgs doublet model motivated by $Z_{2}$ symmetry. The quantum numbers of SM leptons, Higgs doublets,  and vector-like fields are summarized in Table \ref{tab:reps}. A similar model with vector-like quark doublets and singlets was considered in \cite{Dermisek:2019vkc}. While the phenomenology related to vector-like quarks will not be pertinent in this paper, generalizing the model to a 2HDM with a complete VL family is straightforward.

In the basis where the SM lepton Yukawa couplings are diagonal, the most general lagrangian of Yukawa couplings and VL masses under these assumptions is given by

\begin{table}[t]
  \centering
  \begin{threeparttable}
    \begin{tabular}{ccccc|ccc}\label{tab:SM_fields}&\\ \midrule\midrule
    	\makecell{}  & $l_L$ & $e_R$ & $H_u$ & $H_d$ &  $L_{L,R}$ & $N_{L,R}$ & $E_{L,R}$\\
     \cmidrule(l r){1-8}
    \makecell{$SU(2)_L$}  & $\mathbf{2}$ & $\mathbf{1}$ & $\mathbf{2}$ & $\mathbf{2}$ & $\mathbf{2}$ & $\mathbf{1}$ & $\mathbf{1}$ \\
     \cmidrule(l r){1-8}
      \makecell{$U(1)_Y$}  & $-\frac{1}{2}$ & $-1$ & $-\frac{1}{2}$ & $\frac{1}{2}$  & $-\frac{1}{2}$ & $0$ & $-1$\\
           \cmidrule(l r){1-8}
      \makecell{$Z_{2}$}  & $+$ & $-$ & $+$ & $-$  & $+$ & $+$ & $-$\\
      \midrule\midrule
    \end{tabular}
  \end{threeparttable}
   \caption{Quantum numbers of Standard Model leptons, Higgs doublets, and vectorlike leptons under $SU(2)_{L}\times U(1)_{Y}\times Z_{2}$. After electroweak symmetry breaking, the electric charge is given by $Q=T_3 + Y$, where $T_3$ is the weak isospin.}
   \label{tab:reps}
  \end{table}

\begin{flalign}
\label{eq:Lag}
\nonumber
\mathcal{L}\supset& - y_{\mu}\bar{l}_{L}\mu_{R}H_{d} - \lambda_{E}\bar{l}_{L}E_{R}H_{d} - \lambda_{L}\bar{L}_{L}\mu_{R}H_{d} - \lambda\bar{L}_{L}E_{R}H_{d} - \bar{\lambda}H_{d}^{\dagger}\bar{E}_{L}L_{R}\\\nonumber
&-\kappa_{N}\bar{l}_{L}N_{R}H_{u} - \kappa\bar{L}_{L}N_{R}H_{u} -\bar{\kappa}H_{u}^{\dagger}\bar{N}_{L}L_{R}\\
	& - M_{L}\bar{L}_{L}L_{R} - M_{E}\bar{E}_{L}E_{R}  - M_{N}\bar{N}_{L}N_{R}+ h.c.,
\end{flalign}
where the doublet components are labeled as
\begin{equation}
l_{L}=\begin{pmatrix} \nu_{\mu}\\ \mu_{L} \end{pmatrix},\hspace{0.25cm} L_{L,R}=\begin{pmatrix} L_{L,R}^{0}\\ L_{L,R}^{-}\end{pmatrix}, \hspace{0.25cm} H_{d}=\begin{pmatrix} H_{d}^{+}\\ H_{d}^{0}\end{pmatrix},  \hspace{0.25cm} H_{u}=\begin{pmatrix} H_{u}^{0}\\ H_{u}^{-}\end{pmatrix}.
\end{equation}
In the process of electroweak symmetry breaking the neutral components of the Higgs doublets develop vacuum expectation values, $\braket{H^0_{u}}=v_{u}$ and $\braket{H^0_{d}}=v_{d}$, such that $\sqrt{v_u^2 + v_d^2} = v = 174$ GeV, and we define $\tan\beta = v_u/v_d$. Additionally, the charged lepton mass matrix becomes

\begin{equation}
	(\bar{\mu}_{L}, \bar{L}_{L}^{-}, \bar{E}_{L})M_e\begin{pmatrix}\mu_{R}\\ L_{R}^{-}\\ E_{R}\end{pmatrix}=(\bar{\mu}_{L}, \bar{L}_{L}^{-}, \bar{E}_{L})\begin{pmatrix}y_{\mu}v_{d}&0 &\lambda_{E}v_{d}\\\lambda_{L}v_{d}&M_{L}&\lambda v_d\\0&\bar{\lambda}v_{d}& M_{E}\end{pmatrix}\begin{pmatrix}\mu_{R}\\ L_{R}^{-}\\ E_{R}\end{pmatrix}.
\end{equation}
Similarly, for the neutral leptons we obtain
\begin{equation}
	(\bar{\nu}_{\mu}, \bar{L}_{L}^{0}, \bar{N}_{L})M_{\nu}\begin{pmatrix}\nu_{R}=0\\ L_{R}^{0}\\ N_{R}\end{pmatrix}=(\bar{\nu}_{L}, \bar{L}_{L}^{0}, \bar{N}_{L})\begin{pmatrix}0&0 &\kappa_{N}v_{u}\\0&M_{L}&\kappa v_u\\0&\bar{\kappa}v_{u}& M_{N}\end{pmatrix}\begin{pmatrix}\nu_{R}=0\\ L_{R}^{0}\\ N_{R}\end{pmatrix},
\end{equation}
where for convenience we have inserted $\nu_{R}=0$ to present the mass matrix in $3\times3$ form. The mass matrices can be diagonalized by bi-unitary transformations 
\begin{flalign}
\label{eq:mixing_matrices_1}
U_{L}^{e\dagger}&\begin{pmatrix} y_{\mu}v_{d} & 0 & \lambda_{E}v_{d}\\ 
						\lambda_{L}v_{d} & M_{L} & \lambda v_{d}\\
						0 & \bar{\lambda}v_{d}& M_{E}\end{pmatrix}U^{e}_{R}
						=\begin{pmatrix}m_{\mu} & 0 & 0\\
										0 & m_{e_4} & 0\\
										0 & 0 & m_{e_5}
							\end{pmatrix},
\end{flalign}
\begin{flalign}		
\label{eq:mixing_matrices_2}					
U_{L}^{\nu\dagger}&\begin{pmatrix} 0 & 0 & \kappa_{N}v_{u}\\ 
						0 & M_{L} & \kappa v_{u}\\
						0 & \bar{\kappa}v_{u}& M_{N}\end{pmatrix}U^{\nu}_{R}
						=\begin{pmatrix}0 & 0 & 0\\
										0 & m_{\nu_4} & 0\\
										0 & 0 & m_{\nu_5}
							\end{pmatrix},					
\end{flalign}
to obtain lepton mass eigenstates. We label new charged leptons as $e_{4}\text{ and }e_{5}$, and neutral leptons as $\nu_{4}\text{ and }\nu_{5}$. The mixing of VL to the 2nd generation will induce modifications of the muon couplings to gauge and Higgs bosons, leading in particular to flavor non-diagonal lepton couplings. Details of all couplings in the mass eigenstate basis, as well as approximate formulas for individual couplings in the limit of heavy VL masses are given in the Appendix~\ref{sec:Z2_appendix}.

\subsection{2HDM-II-S with vectorlike leptons}
Another well motivated 2HDM-type scenario is the MSSM extended with vectorlike leptons. We do not consider contributions from superpartners which depend on further assumptions about the SUSY-breaking sector. These could be simply added to the contributions from heavy Higgses and VL. Alternatively, our results are complete in the limit of heavy superpartners such that the relevant low-energy particle content of the model is the same as the 2HDM-II-$Z_{2}$. Despite the same particle content, slight differences in the structure of Yukawa couplings will lead to very different results in this case. In the supersymmetric version of the model (2HDM-II-S) the requirement that the superpotential be holomorphic forbids the terms $\bar{\lambda}H_{d}^{\dagger}\bar{E}_{L}L_{R}$ and  $\bar{\kappa}H_{u}^{\dagger}\bar{N}_{L}L_{R}$. However, similar terms are generated through couplings with $H_{u}$ and $H_{d}$ respectively.  We defer to Appendix~\ref{app:SUSY_model} for detailed discussion of the model.

The resulting structure of mixing matrices and couplings follows similarly as in the 2HDM-II-$Z_{2}$ case with the exception that $\bar{\lambda}v_{d}\rightarrow\bar{\lambda}v_{u}$ and $\bar{\kappa}v_{u}\rightarrow\bar{\kappa}v_{d}$ in Eq. \ref{eq:mixing_matrices_1} and \ref{eq:mixing_matrices_2}. This results in replacement of $\bar{\lambda}\rightarrow\bar{\lambda}\tan\beta$ and $\bar{\kappa}\rightarrow\bar{\kappa}/\tan\beta$ in the couplings of gauge bosons and the light SM higgs, while the couplings for the heavy CP-even, CP-odd, and charged Higgses are found with the replacement $\bar{\lambda}\rightarrow-\bar{\lambda}/\tan\beta$ and $\bar{\kappa}\rightarrow-\bar{\kappa}\tan\beta$. In later sections, we will see that this will result in dramatic differences in the predictions for $(g-2)_{\mu}$ compared to the 2HDM-$Z_{2}$ version.

\subsection{SM with vectorlike leptons}
The SM extended with VL and the corresponding contributions to $(g-2)_{\mu}$ have been studied in detail in~\cite{Kannike:2011ng,Dermisek:2013gta}. In section~\ref{sec:results}, we will briefly elaborate on these results, in particular updating the viable parameter space with respect to recent improved measurement of $h\rightarrow \mu^{+}\mu^{-}$. In this case there is essentially no difference in the structure of Yukawa couplings or mixing matrices compared to the 2HDM-II-$Z_{2}$ version of the model with the caveat that the vevs in Eq. \ref{eq:mixing_matrices_1} and \ref{eq:mixing_matrices_2} should be replaced by $v_{d}\rightarrow v$ and $v_{u}\rightarrow v$ (for couplings of the light Higgs this also translates to $\cos\beta\rightarrow 1$ in Eq. \ref{eq:H_couplings} and related approximate formulas).

\section{Contributions to $(g-2)_{\mu}$ from new leptons in two Higgs doublet models}
\label{sec:loops}

\begin{figure}[t]
\centering
\includegraphics[scale=0.75]{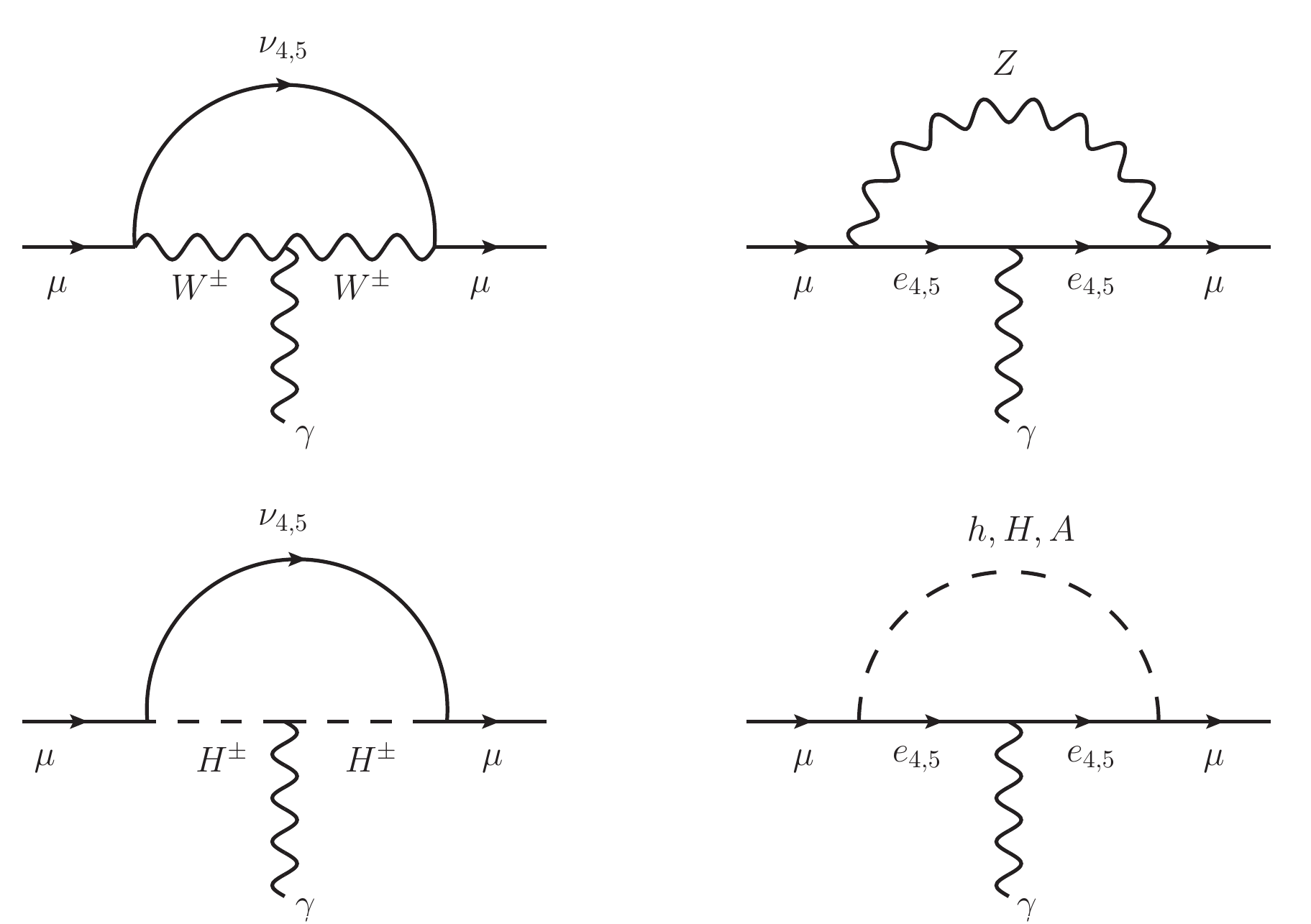}
\caption{Contributions to $(g-2)_{\mu}$ from diagrams involving $W$, $Z$, and Higgs bosons with new leptons.}
\label{fig:diags}
\end{figure}

The 1-loop contributions to $(g-2)_{\mu}$ from new particles induced by mixing with the muon in two Higgs doublet models are shown in Fig.~\ref{fig:diags}. In this section, we present analytical formulas for these contributions in a general two Higgs doublet model. Contributions from SM bosons were previously calculated in~\cite{Kannike:2011ng,Dermisek:2013gta}. 

Defining the couplings of lepton mass eigenstates to the $W$-boson by
\begin{equation}
\mathcal{L}\supset \left(\bar{\hat{\nu}}_{La}\gamma^{\mu}g_{L}^{W\nu_{a}e_{b}}\hat{e}_{Lb}+ \bar{\hat{\nu}}_{Ra}\gamma^{\mu}g_{R}^{W\nu_{a}e_{b}}\hat{e}_{Rb}\right)W^{+}_{\mu} + h.c.,
\end{equation}
the corresponding contribution to $(g-2)_{\mu}$ is
\begin{flalign}
    \Delta a_{\mu}^W = \frac{m_{\mu}}{16 \pi^2 M_W^2}  \mathlarger{\mathlarger{\sum}}_{a = 4, 5} \left[ m_{\mu} \left( |g_R^{W \nu_a \mu} |^2 + |g_L^{W \nu_a \mu}|^2 \right) F_W(x^{a}_{W}) - m_{\nu_a} \text{Re}  \left[ g_R^{W \nu_a \mu } (g_L^{W \nu_a \mu })^{*} \right] G_W(x^{a}_{W}) \right],
\end{flalign}
where $x^{a}_{W}=m_{\nu_{a}}^{2}/M_{W}^{2}$, and the loop functions, $F_W(x)$ and $G_W(x)$, are given by
\begin{flalign}
    F_W(x) =& \frac{4x^4 - 49x^3 + 78x^2 - 43x + 10 + 18 x^3 \ \textrm{ln}(x)}{6(1 - x)^4}, \\
    G_W(x) =&\frac{-x^3 + 12x^2 - 15x + 4 - 6x^2  \text{ln}(x)}{(1 - x)^3}.
\end{flalign}

Similarly, we define couplings of charged or neutral leptons, generically denoted by $f_{a}$, to the $Z$-boson by
\begin{equation}
\mathcal{L}\supset (\bar{f}_{La}\gamma^{\mu}g_{L}^{Zf_{a}f_{b}}f_{Lb} + \bar{f}_{Ra}\gamma^{\mu}g_{R}^{Zf_{a}f_{b}}f_{Rb})Z_{\mu}.
\end{equation}
The $Z$-boson contribution to $(g-2)_{\mu}$ is then given by
\begin{flalign}
    \Delta a_{\mu}^Z = \left( \frac{-m_{\mu}}{8 \pi^2 M_Z^2} \right) \mathlarger{\mathlarger{\sum}}_{a = 4, 5} \left[ m_{\mu} \left( |g_{R}^{Z \mu e_a} |^2 +  |g_{L}^{Z \mu e_a}|^2 \right) F_Z(x^{a}_{Z}) -  m_{e_a} \text{Re} \left[ g_{R}^{Z \mu e_{a}} (g_{L}^{Z \mu e_{a}})^{*} \right] G_Z(x^{a}_{Z}) \right],
\end{flalign}
where the sum is over charged leptons, $e_{4}$ and $e_{5}$, and $x^{a}_{Z}=m_{e_{a}}^{2}/M_{Z}^{2}$. The associated loop functions are given by
\begin{flalign}
    F_Z(x) = &\frac{5x^4 - 14x^3 + 39x^2 - 38x + 8 - 18 x^2 \text{ln}(x)}{12 (1 - x)^4},  \\
    G_Z(x) = &- \frac{x^3 + 3x - 4 - 6x \text{ln}(x)}{2 (1 - x)^3}.
\end{flalign}

Contributions from neutral Higgs bosons $h,H$ and $A$ are identical up to their couplings. For $\phi=h,H,A$ we can define the couplings of charged leptons to neutral Higgses by
\begin{equation}
\mathcal{L}\supset-\frac{1}{\sqrt{2}}\bar{\hat{e}}_{La}\lambda^{\phi}_{e_{a}e_{b}}\hat{e}_{Rb}\phi + h.c.
\end{equation}
The contributions from neutral Higgses to $(g-2)_{\mu}$ involving new charged leptons are then
\begin{flalign}
    \Delta a_{\mu}^{\phi} =   \mathlarger{\mathlarger{\sum}}_{a = 4, 5} \left( \frac{m_{\mu}}{32 \pi^2 m_{\phi}^2} \right) \left[ m_{\mu} \left( |\lambda_{\mu e_a}^{\phi}|^2 + |\lambda_{e_a \mu}^{\phi}|^2 \right) F_{\phi}(x^{a}_{\phi}) + m_{e_a} \text{Re} \left[ \lambda_{\mu e_a}^{\phi} \lambda_{e_a \mu}^{\phi} \right] G_{\phi}(x^{a}_{\phi}) \right],
    \label{eq:Higgs_loops}
\end{flalign}
where $x^{a}_{\phi}=m_{e_{a}}^{2}/m_{\phi}^{2}$ and
\begin{flalign}
    F_{\phi}(x) =& \frac{x^3 - 6x^2 + 3x + 2 + 6x\ \text{ln} (x)}{6 (1 - x)^4}, \\ 
    G_{\phi}(x) =&\frac{-x^2 + 4x - 3 - 2\ \text{ln}(x)}{(1 - x)^3}.
\end{flalign}

Finally, couplings of charged and neutral leptons to the Higgs in the mass eigenstate basis can be defined by
\begin{equation}
\mathcal{L}_{H^{\pm}} = -\bar{\hat{\nu}}_{La}\lambda^{H^{\pm}}_{\nu_{a}e_{b}}\hat{e}_{Rb}H^{+}  - \bar{\hat{e}}_{La}\lambda^{H^{\pm}}_{e_{a}\nu_{b}}\hat{\nu}_{Rb}H^{-} + h.c.
\end{equation}
The contribution to $(g-2)_{\mu}$ from loops with the charged Higgs is then given by
\begin{flalign}
    \Delta a_{\mu}^{H^{\pm}} =  \left( \frac{-m_{\mu}}{16 \pi^2 m_{H^{\pm}}^2} \right) \mathlarger{\mathlarger{\sum}}_{a = 4, 5} \left[ m_{\mu} \left( | \lambda_{\nu_a \mu}^{H^{\pm}}|^2 + |\lambda_{\mu \nu_a}^{H^{\pm}}|^2 \right) F_{H^{\pm}}(x^{a}_{H^{\pm}}) + m_{\nu_a} \text{Re} \left[  \lambda_{\nu_a \mu}^{H^{\pm}} \lambda_{\mu \nu_a}^{H^{\pm}} \right] G_{H^{\pm}}(x^{a}_{H^{\pm}}) \right],
\end{flalign}
where $x^{a}_{H^{\pm}}=m_{\nu_{a}}^{2}/m_{H^{\pm}}^{2}$ and
\begin{flalign}
    F_{H^{\pm}}(x) =& \frac{2x^3 + 3x^2 - 6x + 1 - 6x^2 \ \textrm{ln}(x)}{6 (1 - x)^4}, \\
    G_{H^{\pm}}(x) =& \frac{-x^2 + 1 + 2x \ \textrm{ln}(x)}{ (1 - x)^3}.
\end{flalign}

We emphasize that the formulas given in this section are not specific to any particular 2HDM strucutre (type-I, type-II, type-X, etc.) and can be used in any model with new leptons and extended Higgs sector. Specific contributions to $(g-2)_{\mu}$ in the 2HDM-II-$Z_{2}$ we consider are obtained by inserting the forms of the couplings summarized in the Appendix~\ref{sec:Z2_appendix}. For the 2HDM-II-S and SM appropriate replacements were discussed in the previous section.

Two-loop contributions to $(g-2)_{\mu}$ from Barr-Zee (BZ) diagrams can sometimes be competitive with one-loop predictions due to chiral enhancement in the closed fermion loop, however they are negligible compared to the chirally-enhanced one-loop contributions above. Compared to the one-loop contributions, we find that the typical size of BZ contributions are roughly a factor of $\mathcal{O}(10^{-4} - 10^{-5})$ smaller, see Appendix~\ref{app:BZ} for details. In all the results we present that explain $\Delta a_{\mu}^{exp}$ within 1$\sigma$, the contributions from BZ diagrams are never more than $\mathcal{O}(10^{-3})$ compared to one-loop contributions.

\section{Parameter space and experimental constraints}
\label{sec:constraints}

To study the contributions to $(g-2)_{\mu}$ we vary both dimensionful and dimensionless parameters in the model $\{M_{L},M_{E},M_{N}\}$ and $\{\lambda_{L},\lambda_{E},\lambda,\bar{\lambda},\kappa_{N},\kappa,\bar{\kappa}\}$, $\tan\beta$, and Higgs masses. 

We require $M_{L}>800$ GeV, $M_{E}>200$ GeV, and $M_{N}>100$ GeV in order to generically satisfy constraints from searches for new leptons~\cite{Sirunyan:2019ofn, Aad:2020fzq, Aad:2015dha, Sirunyan:2018mtv}. However, it should be noted that the limits vary significantly with the assumed pattern of branching ratios of new leptons to $W$, $Z$ and $h$~\cite{Dermisek:2014qca} and, in the model we consider an arbitrary pattern of branching ratios can occur~\cite{Dermisek:2015hue} (for a more detailed discussion of branching ratios and approximate formulas for relevant couplings of vectorlike quarks which are completely analogous to leptons, see also Ref.~\cite{Dermisek:2019vkc}). General pattern of branching ratios can allow significantly lighter new leptons than we consider here, especially $SU(2)$ singlets.

For dimensionless parameters we will typically explore values of Yukawa couplings up to $\pm0.5$ or $\pm1$. Values up to $\pm1$ are motivated by perturbativity limits  at very large energy scales, possibly the GUT scale (depending on other details of the model). Occasionally, we will extend the range of couplings up to $\pm\sqrt{4\pi}$ which is motivated by perturbativity limits of couplings at the scale of new physics. 
Note that the signs of three Yukawa couplings are not physical and can be absorbed into a redefinition of three vectorlike lepton fields. For example, $\lambda_L$, $\lambda_E$ and $\kappa_N$ can be chosen to be positive. 

We impose constraints from precision EW data related to the muon and muon neutrino that include $Z$-pole observables, the $W$ partial width, and the muon lifetime. We also impose constraints from oblique corrections \cite{Lavoura:1992np,Chen:2017hak}. These are obtained from data summarized in ref.~\cite{Zyla:2020zbs}. 

  Precision EW measurements constrain possible modification of couplings of the muon to the $Z$ and $W$ bosons at $\sim 0.1\%$ level which, in the limit of small mixing, translates into 95\% C.L. bounds on $\lambda_{E}$ and $\lambda_{L}$ couplings~\cite{Kannike:2011ng}:
\begin{equation}
\Big| \frac{\lambda_E v_d}{M_E}  \Big|   \lesssim 0.03\; , \;\;\;\;\;\; \Big| \frac{\lambda_L v_d}{M_L}  \Big|   \lesssim 0.04
\label{eq:EW_constraints}
\end{equation}
assuming only the Yukawa couplings in the charged sector. In the neutral lepton sector the strongest limits are obtained from the muon lifetime. These were discussed in ref.~\cite{Dermisek:2015oja} together with constraints from 
the invisible widths of the $Z$ boson. The constraint on the $W-\nu-\mu$ coupling translates into an approximate 95\% C.L. upper bound on the size of $\kappa_N$ and $\lambda_E$ couplings:
\begin{equation}
\sqrt{ \left( \frac{\kappa_N v_u}{M_N}\right)^2  + \left( \frac{\lambda_E v_d}{M_E}\right)^2  } \lesssim 0.035,  
\end{equation}
which is slightly lower compared to the one quoted in ref.~\cite{Dermisek:2015oja} due to lower uncertainty in the $W$ mass~\cite{Zyla:2020zbs}.

In  type-II 2HDM neutral Higgs bosons are currently constrained by $H(A) \to t\bar t$ only for $\tan \beta \lesssim 2$~\cite{Sirunyan:2019wph}. At large $\tan \beta$, it is 
the subleading  $H(A) \to \tau^+  \tau^-$ decay mode~\cite{Aad:2020zxo, Sirunyan:2018zut} which leads to stronger limits than  $H(A) \to b\bar b$~\cite{Sirunyan:2018taj, Aad:2019zwb}. Similarly, the strongest limit on the charged Higgs boson at large $\tan \beta$ correspond to the subleading decay mode $H^\pm \to \tau \nu$~\cite{Aaboud:2018gjj,Sirunyan:2019hkq}. The limits on $H^+ \to t\bar b$ are currently weaker at  large $\tan \beta$, however they also constrain charged Higgs masses below $\tan \beta \simeq 2$~\cite{Aaboud:2018cwk, Sirunyan:2019arl, ATLAS:2021upq}.

For simplicity, for the 2HDM-II-$Z_{2}$ we assume degenerate heavy Higgs masses $m_{A}=m_{H}=m_{H^{\pm}}$ and for 2HDM-II-S we assume the standard tree-level relations between masses of heavy Higgs bosons. Thus, we only impose ATLAS limits on  $H(A) \to \tau^+  \tau^-$~\cite{Aad:2020zxo} and on $H^+ \to t\bar b$~\cite{ATLAS:2021upq} which are currently the strongest at large and small $\tan \beta$ respectively. These assumptions are also sufficient to satisfy constraints from flavor observables~\cite{Haller:2018nnx}.

In addition to constraints on heavy Higgs masses, there are relevant constraints on the SM Higgs coupling to the muon through its modified relation to the muon mass. In the present case, the physical muon mass originates from its coupling to $H_{d}$ as well as mixing with heavy leptons
\begin{equation}
m_{\mu}\simeq y_{\mu}v_{d} + m_{\mu}^{LE},
\label{eq:muon_mass}
\end{equation}
where we have defined
\begin{equation}
m_{\mu}^{LE}\equiv \frac{\bar{\lambda}\lambda_{L}\lambda _{E}  v^{3}\cos^{3}\beta }{M_{E} M_{L}},
\label{eq:mLE}
\end{equation}
that would give the muon mass in the absence of $y_{\mu}$ as can be seen from the determinant of Eq. \ref{eq:mixing_matrices_1}. Thus, for a given set of parameters that fix $m_{\mu}^{LE}$, $y_{\mu}$ can be iteratively determined so that Eq. \ref{eq:muon_mass} leads to the measured value of the muon mass. However, the sign of the muon mass determined by Eq. \ref{eq:muon_mass} is not physical and thus there are two solutions, $y^{\pm}_{\mu}$, leading to $\pm m_{\mu}$, either of which is acceptable in principle. The wrong-sign of the mass can always be rotated away by proper field redefinition of eigenstates. Due to the arbitrary overall sign of $m_{\mu}^{LE}$ it is always possible to restrict to $y^{+}_{\mu}$ solutions.

From the Higgs coupling to the muon 
\begin{equation}
\lambda^{h}_{\mu\mu}\simeq y_{\mu}\cos\beta + 3m^{LE}_{\mu}/v \simeq (m_{\mu} + 2m_{\mu}^{LE})/v,
\label{eq:Higgs_muon_coupling}
\end{equation}
it follows that $\lambda^{h}_{\mu\mu}> 3(\lambda^{h}_{\mu\mu})^{SM}$ when $m_{\mu}^{LE}>m_{\mu}$. Current measurements of the $h\rightarrow \mu^{+}\mu^{-}$ decay~\cite{Aad:2020xfq} by far exclude this possibility. Thus, in our numerical analysis we restrict to regions of parameters where $m_{\mu}^{LE}<m_{\mu}$, and thus $y^{+}_{\mu}>0$. We will explore the impact of $h\rightarrow \mu^{+}\mu^{-}$ constraints in this region further in the following section.

We note that similar loops as in Fig. \ref{fig:diags} will also generate a correction to the muon Yukawa coupling. This could lead to large corrections to $y_{\mu}$ compared to the value needed to reproduce the muon mass. As a simple example, we will see in the following sections that regions of parameters which achieve $\Delta a_{\mu}^{exp}$ within 1$\sigma$ in the SM also require that the tree-level Higgs coupling to the muon is typically $y_{\mu}\simeq 2m_{\mu}/v$. Loop corrections to the muon Yukawa coupling in our model scale as $\Delta y_{\mu}\simeq\lambda_{L}\lambda_{E}\bar{\lambda}/8\pi^{2}$ and reach this value for couplings~$\sim0.5$. For couplings of order 1, motivated by perturbitivity in the UV, a tuning of only about 10\% between tree- and loop-level contributions to $y_{\mu}$ is expected in these scenarios. However, it could be argued that scenarios with larger couplings suffer from a fine-tuning problem with respect to the physical muon mass. See also~\cite{Capdevilla:2021rwo} for a related discussion.

\section{Results}
\label{sec:results}
The current measurement of the muon anomalous magnetic moment sits at more than four standard deviations from the predicted value in the SM~\cite{Abi:2021gix, Aoyama:2020ynm}
\begin{equation}
\Delta a^{exp}_{\mu}\equiv a^{exp}_{\mu} - a_{\mu}^{SM} = (2.51 \pm 0.59)\times 10^{-9}.
\end{equation}
Contributions to $\Delta a_{\mu}$ from charged and neutral vectorlike leptons with mixing to the muon are given by loops with $h, Z$ and $W$ bosons as well as those with heavy Higgses, $A, H$ and $H^{\pm}$. The contributions involving vectorlike leptons and SM bosons were calculated previously in \cite{Kannike:2011ng,Dermisek:2013gta}. The complementarity of contributions from charged vectorlike leptons to $\Delta a_{\mu}$ and other precision observables in a 2HDM-II-$Z_{2}$ was presented in \cite{Dermisek:2020cod}. In this paper, we extend the calculation to include mixing in the neutral lepton sector. In the following subsections, we provide a detailed study of the 2HDM-II-$Z_{2}$ followed by a discussion of the corresponding predictions for $\Delta a_{\mu}$ in the 2HDM-II-S. We also compare these results to the current status of the SM with VL.

\subsection{2HDM-II-$Z_{2}$ with vectorlike leptons}
Contributions to $(g-2)_{\mu}$ can be calculated following the analytic formulas in section~\ref{sec:loops} and Appendix~\ref{sec:Z2_appendix}. In the following, it will prove useful to have approximations on hand to estimate the impact of individual particles to $\Delta a_{\mu}$ in terms of lagrangian parameters. In Tables~\ref{gm2_leading_doublet} and \ref{gm2_leading_singlet} we summarize individual contributions from doublet- and singlet-like new leptons to $\Delta a_{\mu}$ normalized by $m_{\mu}/16\pi^2$, in the limit of VL masses well above the EW scale (note the comments after Eq.~\ref{eq:A64} for the appropriate approximations used). We also assume that the masses of heavy Higgs bosons are comparable to that of new leptons.\footnote{Our approximations are accurate to within 10\% in the range $\frac{1}{\sqrt{2}}M_{L,E,N} \lesssim m_{H,A,H^{\pm}} \lesssim \sqrt{2}M_{L,E,N}$. Though, in our numerical results we do not use any approximations. For heavier lepton masses, $M_{L,E,N}\gg m_{H,A,H^{\pm}}$, one can make the replacements $\frac{1}{6}\left(\frac{M_{L,N}^{2}}{m_{H^{\pm}}^{2}} + 1\right)\rightarrow 1$ and $\left(\frac{M_{L,E}^{2}}{6m_{H,A}^{2}} + \frac{1}{2}\right)\rightarrow 1$ in Tables~\ref{gm2_leading_doublet} and \ref{gm2_leading_singlet} for charged and neutral Higgs contributions, respectively. In the opposite limit, $M_{L,E,N}\ll m_{H,A,H^{\pm}}$, the corresponding replacements are $\frac{1}{6}\left(\frac{M_{L,N}^{2}}{m_{H^{\pm}}^{2}} + 1\right)\rightarrow \frac{M_{L,N}^{2}}{m_{H^{\pm}}^{2}}$ and $\left(\frac{M_{L,E}^{2}}{6m_{H,A}^{2}} + \frac{1}{2}\right)\rightarrow (-3-2\ln(M_{L,E}^{2}/m_{H,A}^{2}))(M_{L,E}^{2}/m_{H,A}^{2})$. We note that the latter expansion for neutral Higgs loops is numerically good to within a factor of 2 up to $x_{a}\simeq 0.1$.} The derivation of each contribution is straightforward from approximate couplings listed in Appendix~\ref{approx_formulas}. The total approximate contributions assuming heavy lepton masses can be found by summing the corresponding rows in the tables. We find
\begin{table}[t]
\centering
\begin{tabular}{|c|c|}
	\hline
	$\frac{16\pi^{2}}{m_{\mu}}\Delta a^{i}_{\mu} $  & $SU(2)$ doublets \\
	\hline\hline
	$Z$ & $\frac{1}{2M_{E}}\frac{\lambda_{L}v_{d}\cos^{2}\beta}{M_{E}^{2}-M_{L}^{2}}\left[\lambda_{E}\bar{\lambda}M_{L} + \lambda_{E}\lambda M_{E}\right]$ \\
	\hline
	$W$ & $\frac{\lambda_{L}\cos\beta}{M_{L}}\left[ \frac{v_{d}}{M_{E}}\lambda_{E}\bar{\lambda}\cos\beta - \frac{v_{u}\sin\beta}{M_{N}^{2}-M_{L}^{2}}\left(\kappa_{N}\bar{\kappa}M_{N} + \kappa_{N}\kappa M_{L}\right)\right]$  \\
	\hline
	$H^{\pm}$ &  $\frac{1}{6}\left(\frac{M_{L}^{2}}{m_{H^{\pm}}^{2}}  + 1\right)\frac{\lambda_{L}}{M_{L}}\left[\frac{v_{d}}{M_{E}}\lambda_{E}\bar{\lambda}\sin^2\beta + \frac{v_{u}\sin\beta\cos\beta}{M_{N}^{2} - M_{L}^{2}}\left(\kappa_{N}\bar{\kappa}M_{N} + \kappa_{N}\kappa M_{L}\right)\right]$  \\
	\hline
	$h$ & $-\frac{1}{2M_{L}}\lambda_{L}\cos\beta\left[\frac{v_{d}}{M_{E}}\lambda_{E}\bar{\lambda}\cos\beta + \frac{v_{d}\cos\beta}{M_{E}^{2}-M_{L}^{2}}\left(\lambda_{E}\bar{\lambda}M_{E} + \lambda_{E}\lambda M_{L}\right)\right]$\\
	\hline
	$H$ & $-\frac{1}{2}\left(\frac{M_{L}^2}{6m_{H}^{2}} + \frac{1}{2}\right)\frac{\lambda_{L}\sin\beta}{M_{L}}\left[\frac{v_{d}}{M_{E}}\lambda_{E}\bar{\lambda}\sin\beta + \frac{v_{d}\sin\beta}{M_{E}^{2}-M_{L}^{2}}\left(\lambda_{E}\bar{\lambda}M_{E} + \lambda_{E}\lambda M_{L}\right)\right]$  \\
	\hline
	$A$ & $-\frac{1}{2}\left(\frac{M_{L}^2}{6m_{A}^{2}} + \frac{1}{2}\right)\frac{\lambda_{L}\sin\beta}{M_{L}}\left[\frac{v_{d}}{M_{E}}\lambda_{E}\bar{\lambda}\sin\beta - \frac{v_{d}\sin\beta}{M_{E}^{2}-M_{L}^{2}}\left(\lambda_{E}\bar{\lambda}M_{E} + \lambda_{E}\lambda M_{L}\right)\right]$\\
	\hline
\end{tabular}
\caption{Leading contributions to $\Delta a^{i}_{\mu}$ from new lepton doublets assuming $m_{A,H,H^{\pm}}\simeq M_{L,E,N}\gg M_{Z}$.}
\label{gm2_leading_doublet}
\end{table}
\begin{table}[t]
\centering
\begin{tabular}{|c|c|}
	\hline
	$\frac{16\pi^{2}}{m_{\mu}}\Delta a^{i}_{\mu} $  & $SU(2)$ singlets \\
	\hline\hline
	$Z$ &  $-\frac{1}{2M_{L}}\frac{\lambda_{E}v_{d}\cos^{2}\beta}{M_{E}^{2}-M_{L}^{2}}\left[\lambda_{L}\bar{\lambda}M_{E} + \lambda_{L}\lambda M_{L}\right]$ \\
	\hline
	$W$ &  $\frac{\kappa_{N}\sin\beta}{M_{L}}\left[\frac{v_{d}\sin\beta}{M_{N}^{2}-M_{L}^{2}}\left(\lambda_{L}\kappa M_{L} + \lambda_{L}\bar{\kappa} M_{N}\right)\right]$  \\
	\hline
	$H^{\pm}$ &  $\frac{1}{6}\left(\frac{M_{N}^{2}}{m_{H^{\pm}}^{2}}  + 1\right)\frac{\kappa_{N}}{M_{N}}\cos\beta\left[\frac{v_{d}}{M_{L}}\lambda_{L}\bar{\kappa}\cos\beta - \frac{v_{u}\sin\beta}{M_{N}^{2}-M_{L}^{2}}\left(\lambda_{L}\bar{\kappa}M_{L} + \lambda_{L}\kappa M_{N}\right)\right]$ \\
	\hline
	$h$ & $-\frac{1}{2M_{E}}\lambda_{E}\cos\beta\left[\frac{v_{d}}{M_{L}}\lambda_{L}\bar{\lambda}\cos\beta - \frac{v_{d}\cos\beta}{M_{E}^{2}-M_{L}^{2}}\left(\lambda_{L}\bar{\lambda}M_{L} + \lambda_{L}\lambda M_{E}\right)\right]$\\
	\hline
	$H$ &  $-\frac{1}{2}\left(\frac{M_{E}^{2}}{6m_{H}^{2}} + \frac{1}{2}\right)\frac{\lambda_{E}\sin\beta}{M_{E}}\left[\frac{v_{d}}{M_{L}}\lambda_{L}\bar{\lambda}\sin\beta - \frac{v_{d}\sin\beta}{M_{E}^{2}-M_{L}^{2}}\left(\lambda_{L}\bar{\lambda}M_{L} + \lambda_{L}\lambda M_{E}\right)\right]$\\
	\hline
	$A$ & $-\frac{1}{2}\left(\frac{M_{E}^{2}}{6m_{A}^{2}} + \frac{1}{2}\right)\frac{\lambda_{E}\sin\beta}{M_{E}}\left[\frac{v_{d}}{M_{L}}\lambda_{L}\bar{\lambda}\sin\beta + \frac{v_{d}\sin\beta}{M_{E}^{2}-M_{L}^{2}}\left(\lambda_{L}\bar{\lambda}M_{L} + \lambda_{L}\lambda M_{E}\right)\right]$ \\
	\hline
\end{tabular}
\caption{Leading contributions to $\Delta a^{i}_{\mu}$ from new lepton singlets assuming $m_{A,H,H^{\pm}}\simeq M_{L,E,N}\gg M_{Z}$.}
\label{gm2_leading_singlet}
\end{table}
\begin{equation}
\label{eq:Z_contribution}
\Delta a_{\mu}^{Z} \simeq -\frac{m_{\mu}vc^{3}_{\beta}}{32\pi^{2}}\frac{\lambda_{L}\lambda_{E}\bar{\lambda}}{M_{L}M_{E}},
\end{equation}
\begin{equation}
\label{eq:W_contribution}
\Delta a_{\mu}^{W} \simeq \frac{m_{\mu}vc^{3}_{\beta}}{16\pi^{2}}\frac{\lambda_{L}\lambda_{E}\bar{\lambda}}{M_{E}M_{L}},
\end{equation}
\begin{equation}
\label{eq:h_contribution}
\Delta a_{\mu}^{h} \simeq -\frac{3m_{\mu}vc^{3}_{\beta}}{32\pi^{2}}\frac{\lambda_{L}\lambda_{E}\bar{\lambda}}{M_{E}M_{L}},
\end{equation}
for gauge bosons and SM-like Higgs. For the contributions from heavy Higgses with masses comparable to new leptons we find 
\begin{flalign}
\label{eq:Hpm_contribution}\nonumber
\Delta a_{\mu}^{H^{\pm}} \simeq \frac{m_{\mu}vs^{2}_{\beta}c_{\beta}}{96\pi^{2}m_{H^{\pm}}^{2}}\Bigg[&\left(\frac{\lambda_{L}\lambda_{E}\bar{\lambda}M_{L}}{M_{E}} + \frac{\lambda_{L}\kappa_{N}\bar{\kappa}M_{N}}{M_{L}\tan^{2}\beta} - \lambda_{L}\kappa_{N}\kappa\right)\\
& + \frac{\lambda_{L}\lambda_{E}\bar{\lambda}m_{H^{\pm}}^{2}}{M_{L}M_{E}} + \frac{\lambda_{L}\kappa_{N}\bar{\kappa}m_{H^{\pm}}^{2}}{M_{L}M_{N}}\left(\frac{1}{\tan^{2}\beta} + 1\right)\Bigg],
\end{flalign}
\begin{equation}
\label{eq:H_contribution}
\Delta a_{\mu}^{H} \simeq- \frac{m_{\mu}vs^{2}_{\beta}c_{\beta}}{192\pi^{2}m_{H}^{2}}\left[\left(\frac{\lambda_{L}\lambda_{E}\bar{\lambda}(M_{L}^{2}+M_{E}^{2})}{M_{L}M_{E}} - \lambda_{L}\lambda_{E}\lambda\right) + 9m_{H}^{2}\frac{\lambda_{L}\lambda_{E}\bar{\lambda}}{M_{L}M_{E}}\right],
\end{equation}
\begin{equation}
\label{eq:A_contribution}
\Delta a_{\mu}^{A} \simeq -\frac{m_{\mu}vs^{2}_{\beta}c_{\beta}}{192\pi^{2}m_{A}^{2}}\left[\left(\frac{\lambda_{L}\lambda_{E}\bar{\lambda}(M_{L}^{2}+M_{E}^{2})}{M_{L}M_{E}} + \lambda_{L}\lambda_{E}\lambda\right) + 3m_{A}^{2}\frac{\lambda_{L}\lambda_{E}\bar{\lambda}}{M_{L}M_{E}}\right].
\end{equation}
These equations are also valid when $M_{L}=M_{E}=M_{N}$ unlike the approximations of separate contributions in Tables~\ref{gm2_leading_doublet} and \ref{gm2_leading_singlet}.
They further simplify when all up-type couplings are zero and masses of all new particles are equal $M_{L,E} = m_{H,A,H^\pm}$. In this limit, the contributions can be parameterized as $\Delta a^{i}_{\mu} \simeq   \frac{k^i}{16\pi^{2}}  \frac{m_\mu m_\mu^{LE}}{v^2}$, where $k^W =   1$, $k^Z =   -1/2$, $k^h = -3/2$, $k^H = - (11/12) \tan^2 \beta$, $k^A = - (5/12) \tan^2 \beta$, and $k^{H^\pm} = (1/3) \tan^2 \beta$~\cite{Dermisek:2020cod}. We have additionally ignored terms $\propto \lambda$ in the CP-even and CP-odd Higgs contributions as these terms would cancel in the total contribution when $m_{H}=m_{A}$. Note that $k^W + k^Z + k^h = -1$, while $k^H + k^A + k^{H^\pm} = - \tan^2 \beta$.
\begin{figure}[t]
\centering
\includegraphics[scale=0.23]{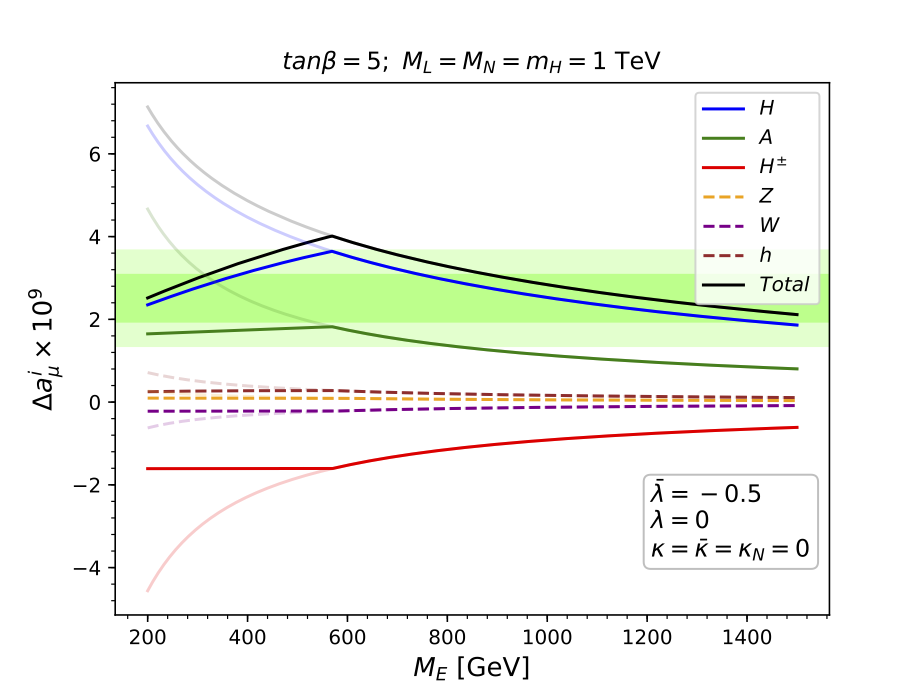}
\includegraphics[scale=0.23]{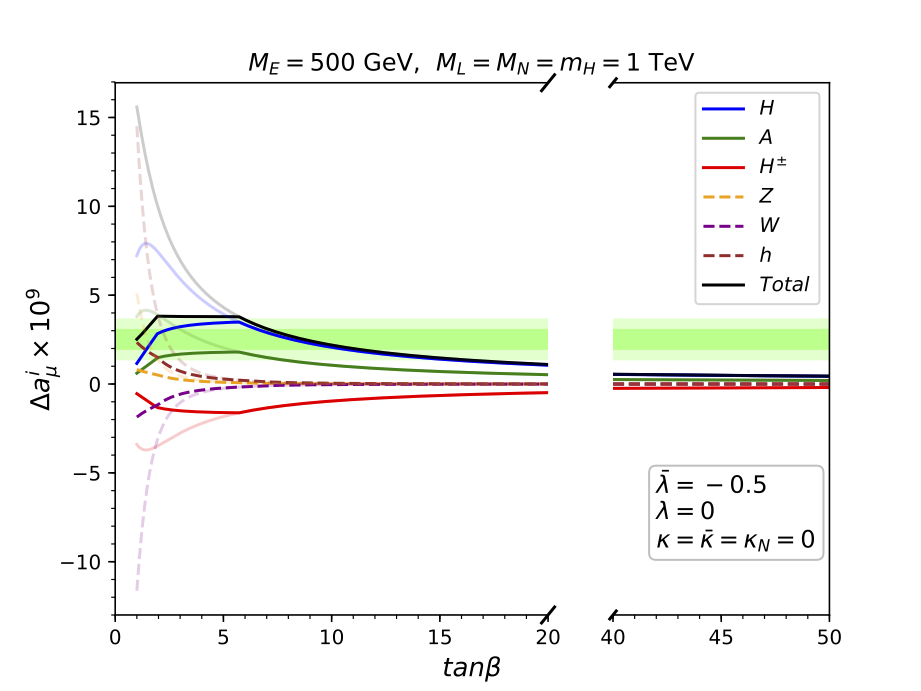}
\caption{Individual contributions to $\Delta a_{\mu}\times10^{9}$ as a function of $M_{E}$ for $\tan\beta = 5$ (Left) and as a function of $\tan\beta$ for $M_{E} = 500$ GeV (Right) with $M_{L,N}=m_{H}=1$ TeV. The total contribution is shown with the solid black curve. Shaded curves show the behavior when $\lambda_{L}=\lambda_{E}=0.5$, while solid curves show the contributions when $\lambda_{L}$ and $\lambda_{E}$ are fixed to the maximum values allowed by precision EW constraints up to $0.5$. Other couplings are fixed to $\bar{\lambda}=-0.5$, $\lambda=\kappa=\bar{\kappa}=\kappa_{N}=0$. The dark and light shaded green bands represent the 1 and 2$\sigma$ levels of $\Delta a_{\mu}^{exp}$, respectively.}
\label{fig:line_plot}
\end{figure}

We see that the leading contribution from SM bosons is  $\propto - m_{\mu}^{LE}$ and likewise for CP-even and CP-odd heavy Higgses up to terms proportional to $\lambda$. The terms in the charged Higgs contribution proportional to $m_{\mu}^{LE}$ are positive leading to possible cancelation with other contributions. However, due to mixing in the neutral lepton sector the charged Higgs loop receives additional contributions which can independently control the overall sign. Further, when $m_{\mu}^{LE}=0$ the charged Higgs loop gives the dominant contribution from new physics to $(g-2)_{\mu}$ in the leading order approximation assuming comparable masses of new leptons.

The behavior of individual loop contributions to $\Delta a_{\mu}$ with respect to $M_{E}$ and $\tan\beta$ are shown in Fig.~\ref{fig:line_plot} for a representative choice of parameters. In both panels, $M_{L}=M_{N}=m_{H}=1$ TeV are fixed. The solid-color curves correspond to scenarios when these couplings are fixed to their maximum values allowed by EW precision constraints up to values of 0.5. Ignoring EW precision constraints and instead fixing  $\lambda_{L}=\lambda_{E}=0.5$ individual contributions follow the corresponding shaded curves. For simplicity we also fix $\bar{\lambda}=-0.5$. The signs are chosen to illustrate the impact from positive contributions of $H$ and $A$. Opposite signs of contributions would be found if the sign of $\lambda_{L}$, $\lambda_{E}$, or $\bar{\lambda}$ were flipped. Other couplings are fixed to $\lambda=\kappa=\bar{\kappa}=\kappa_{N}=0$ for simplicity. 

The kinks seen in the curves occur when the precision EW constraints, Eq (\ref{eq:EW_constraints}), become saturated, $\lambda_{L}=0.04M_{L}/v_{d}$ and $\lambda_{E}=0.03M_{E}/v_{d}$. In the left plot, contributions from SM bosons are independent of $M_{E}$ below the kink, while contributions from heavy neutral Higgses increase with terms proportional to $M_{E}^{2}$. Similar contributions from charged Higgs scale as $M_{L}^{2}$. For $M_{E}$ above the kink, all contributions asymptote to zero as heavy particles are decoupled. Note that the range of masses which can explain $\Delta a_{\mu}^{exp}$ is highly sensitive to $\tan\beta$ where larger $\tan\beta$ requires either larger couplings or lower masses. 
\begin{figure}[t]
\centering
\includegraphics[scale=0.23]{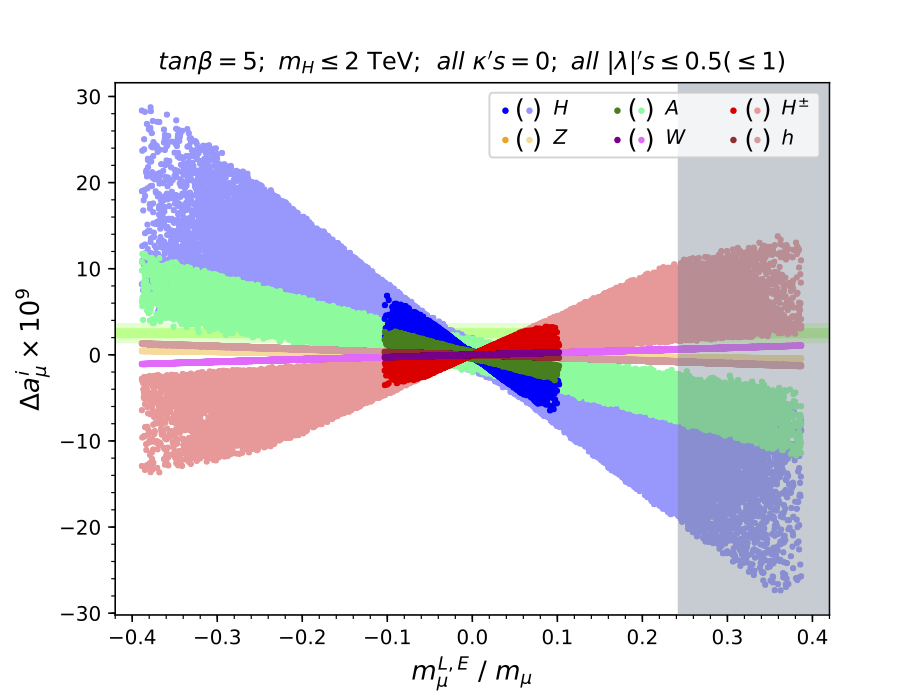}
\includegraphics[scale=0.23]{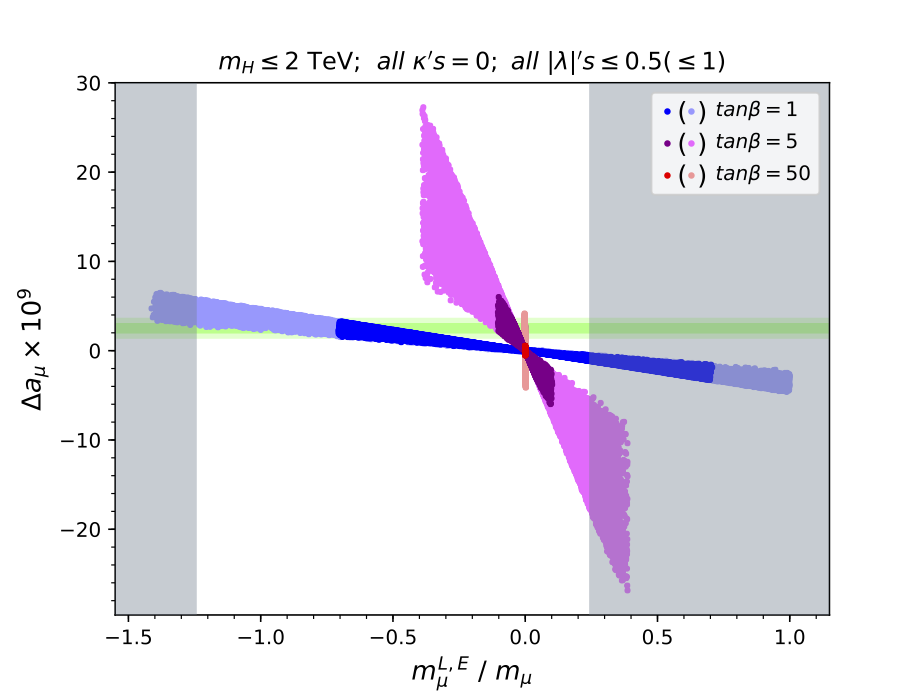}
\caption{Individual contributions to $(g-2)_{\mu}$ with respect to $m_{\mu}^{LE}/m_{\mu}$ for $\tan\beta = $5 (left) and total contributions for $\tan\beta=1,5,50$ (right) assuming down-type couplings up to 0.5 (dark colors) and 1 (shaded colors). For simplicity we fix $\kappa=\bar{\kappa}=\kappa_{N}=0$. $m_{H}=m_{A}=m_{H^{\pm}}$ are scanned up to 2 TeV subject to all constraints discussed in section~\ref{sec:constraints}. Gray shaded regions are ruled out by $h\rightarrow \mu^{+}\mu^{-}$. The dark and light shaded green bands represent the 1 and 2$\sigma$ levels of $\Delta a_{\mu}^{exp}$, respectively.}
\label{fig:scan_plot}
\end{figure}
In Fig.~\ref{fig:scan_plot}, we the show the range of individual (left) and total (right) contributions to $\Delta a_{\mu}$ with respect to $m_{\mu}^{LE}/m_{\mu}$ for couplings up to 0.5 (dark colors) and 1 (shaded colors). All up-type couplings are fixed $\kappa=\bar{\kappa}=\kappa_{N}=0$ (scanning over $\kappa$'s would give almost identical results). From the individual contributions for $\tan\beta=5$ in the left panel, we see that due to the $\tan^{2}\beta$ enhancement, heavy Higgs contributions can give an order of magnitude larger contribution than those from SM bosons over most of the parameter space. For instance, for comparable heavy lepton and Higgs masses $\Delta a_{\mu}^{H}\simeq 0.6\tan^{2}\beta\times\Delta a_{\mu}^{h}$. Note that since CP-odd and charged Higgs contributions tend to cancel the enhancement is largely driven by the CP-even Higgs contribution for most of the parameter space. This can be seen when comparing to the right panel of Fig.~\ref{fig:scan_plot} where we show the total contribution to $(g-2)_{\mu}$ for $\tan\beta=1,5,50$. In both panels we show the regions of $m_{\mu}^{LE}/m_{\mu}$ that are excluded by $h\rightarrow \mu^{+}\mu^{-}$. Note that both $m_{\mu}^{LE}/m_{\mu}=0$ and -1 lead to the same prediction of $h\rightarrow \mu^{+}\mu^{-}$ as in the SM which can be seen from Eq. (\ref{eq:Higgs_muon_coupling}).

In both panels, the dark and light shaded green bands represent the 1 and $2\sigma$ levels of $\Delta a_{\mu}^{exp}$, respectively. We see that for couplings up to 0.5 (1), the correction to the magnetic moment spans a range about 4 (10) times the measured central value. As a curiosity, we mention that allowing couplings up to the perturbativity limit, $\sim\sqrt{4\pi}$, the possible contribution to $\Delta a_{\mu}\sim 200\times 10^{-9}$ can be achieved while still satisfying all relevant constraints.

Regarding contributions from up-type couplings, it is clear from Tables \ref{gm2_leading_doublet} and \ref{gm2_leading_singlet} that corrections to $(g-2)_{\mu}$ from charged currents are the only relevant pieces. Mixing with the neutral component of the doublets $L_{L,R}^{0}$ further dictates that additionally $\lambda_{L}$ should be non-zero to have any non-vanishing effects from $\kappa$'s at leading order. Further, it is expected that any effect from loops in involving the $W$-boson are small, see Eq.~\ref{eq:W_contribution}, since the leading order contributions from $SU(2)$ doublets tend to cancel those from singlets. In fact, we find that the sub-leading contribution from the $W$-loop can be found by further expanding $x_{a}G_{W}(x_{a})$ at the next order in $x_{a}$
\begin{flalign}\nonumber
\Delta a_{\mu}^{W}\simeq\;&\frac{6m_{\mu}M_{W}^{2}v}{16\pi^{2}}\frac{\lambda_{L}\kappa_{N}s^{2}_{\beta}c_{\beta}}{M_{N}^{2}M_{L}^{2}}\left(\bar{\kappa}\frac{M_{N}}{M_{L}}+\kappa\right)\\
&\Bigg[\frac{M_{L}^{2}}{M_{N}^{2}-M_{L}^{2}}\ln\left(M_{N}^{2}/M_{W}^{2}\right)-\frac{M_{N}^{2}}{M_{N}^{2}-M_{L}^{2}}\ln\left(M_{L}^{2}/M_{W}^{2}\right) + \frac{3}{2}\Bigg].
\label{eq:W_sub}
\end{flalign}
\begin{figure}[t]
\centering
\includegraphics[scale=0.23]{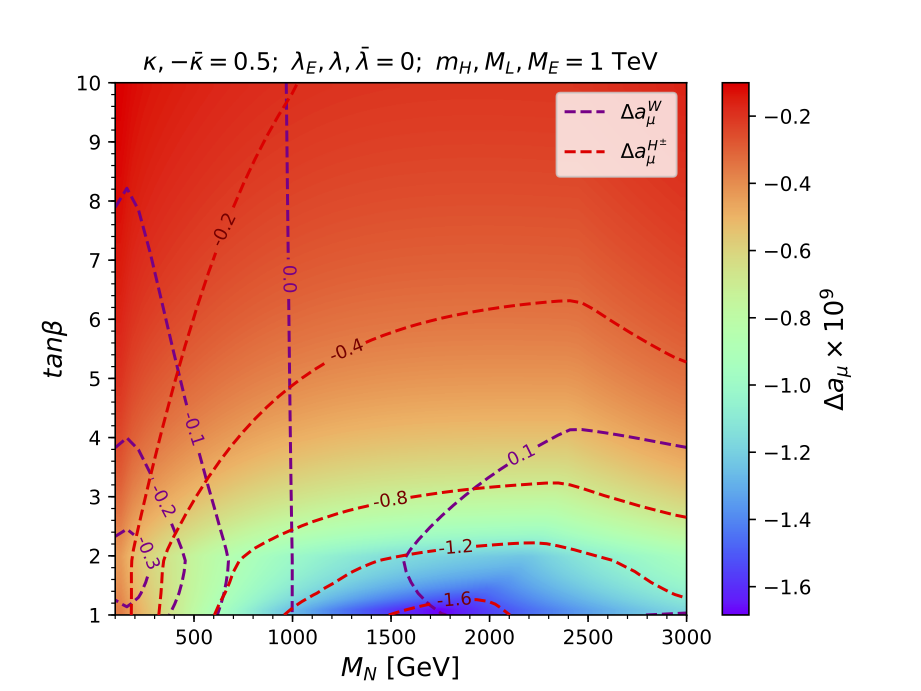}
\includegraphics[scale=0.23]{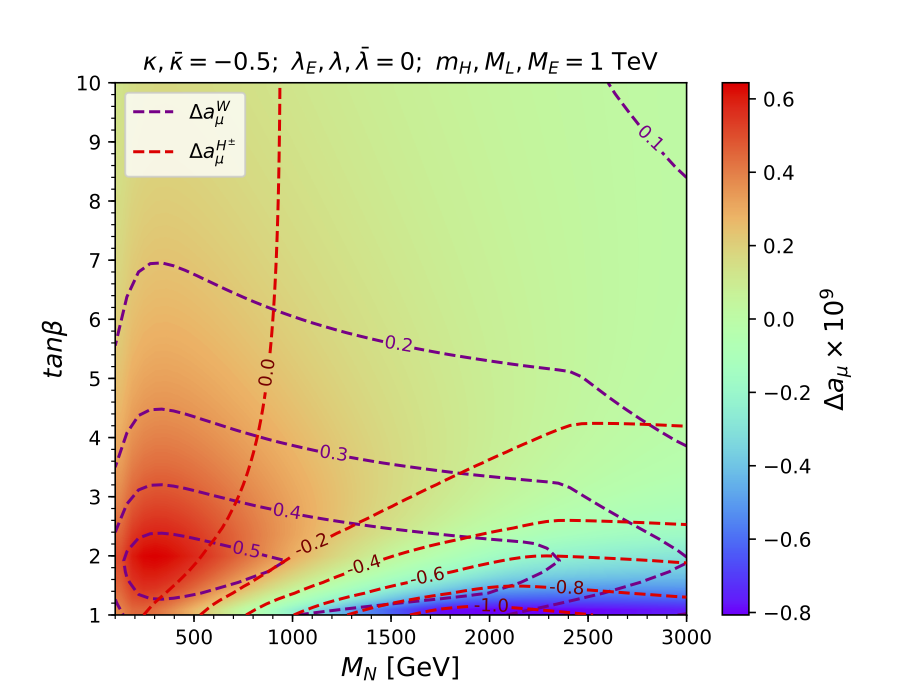}
\caption{{\textbf{Left:}} Predictions for $\Delta a_{\mu}$ with respect to $M_{N}$ and $\tan\beta$ when only $\lambda_{L},\kappa,\bar{\kappa}$, and $\kappa_{N}$ are non-zero. $\kappa=-\bar{\kappa}=0.5$ are fixed and both $\lambda_{L}$ and $\kappa_{N}$ are set to their maximum values allowed by precision EW constraints. Purple and red dashed curves show individual contributions from $W$ and $H^{\pm}$ loops respectively. {\textbf{Right:}} Same contours as in the left panel with $\kappa=\bar{\kappa}=-0.5$.}
\label{fig:kappa_plot}
\end{figure}
In Fig.~\ref{fig:kappa_plot} we show the size of corrections to $(g-2)_{\mu}$ (color shading) in the limit that only $\lambda_{L},\kappa,\bar{\kappa}$, and $\kappa_{N}$ are non-zero. We have fixed $\lambda_{L}$ and $\kappa_{N}$ to their maximum values allowed by precision EW constraints. In the left panel we fix $\kappa=-\bar{\kappa}=0.5$, while in the right panel both couplings are chosen to have the same sign. We explore both cases of the relative sign since individual contributions are sensitive to this choice. The $W$ contribution, shown in purple dashed curves, can switch signs depending on whether $\nu_{4}$ is mostly singlet- or doublet-like, dictated by the prefactor $\left(\bar{\kappa}\frac{M_{N}}{M_{L}}+\kappa\right)$ in Eq. (\ref{eq:W_sub}). However, the charged Higgs contribution (dashed red curves) remains negative in the entire plane (note that opposite sign of individual contributions shown is also possible simply by replacing $\lambda_{L}\rightarrow - \lambda_{L}$). 

We see that for small $\tan\beta$ and relatively large $M_{N}$ the charged Higgs contribution can alone explain $\Delta a_{\mu}^{exp}$ within 2$\sigma$ for couplings up to 0.5 (when $\kappa$ and $\bar{\kappa}$ have opposite sign). Allowing couplings up to 1, the charged Higgs contribution could even explain the central value of $\Delta a_{\mu}^{exp}$ in this region of parameters. It should be noted that the size of contributions shown in the plane are completely orthogonal to contributions resulting from $m_{\mu}^{LE}$ being non-zero. Thus, one can simply add the size of contributions from the previous figures to Fig.~\ref{fig:kappa_plot} to estimate the total contribution to $(g-2)_{\mu}$ for a given choice of masses and $\tan\beta$.
\begin{figure}[h]
\centering
\includegraphics[scale=0.47]{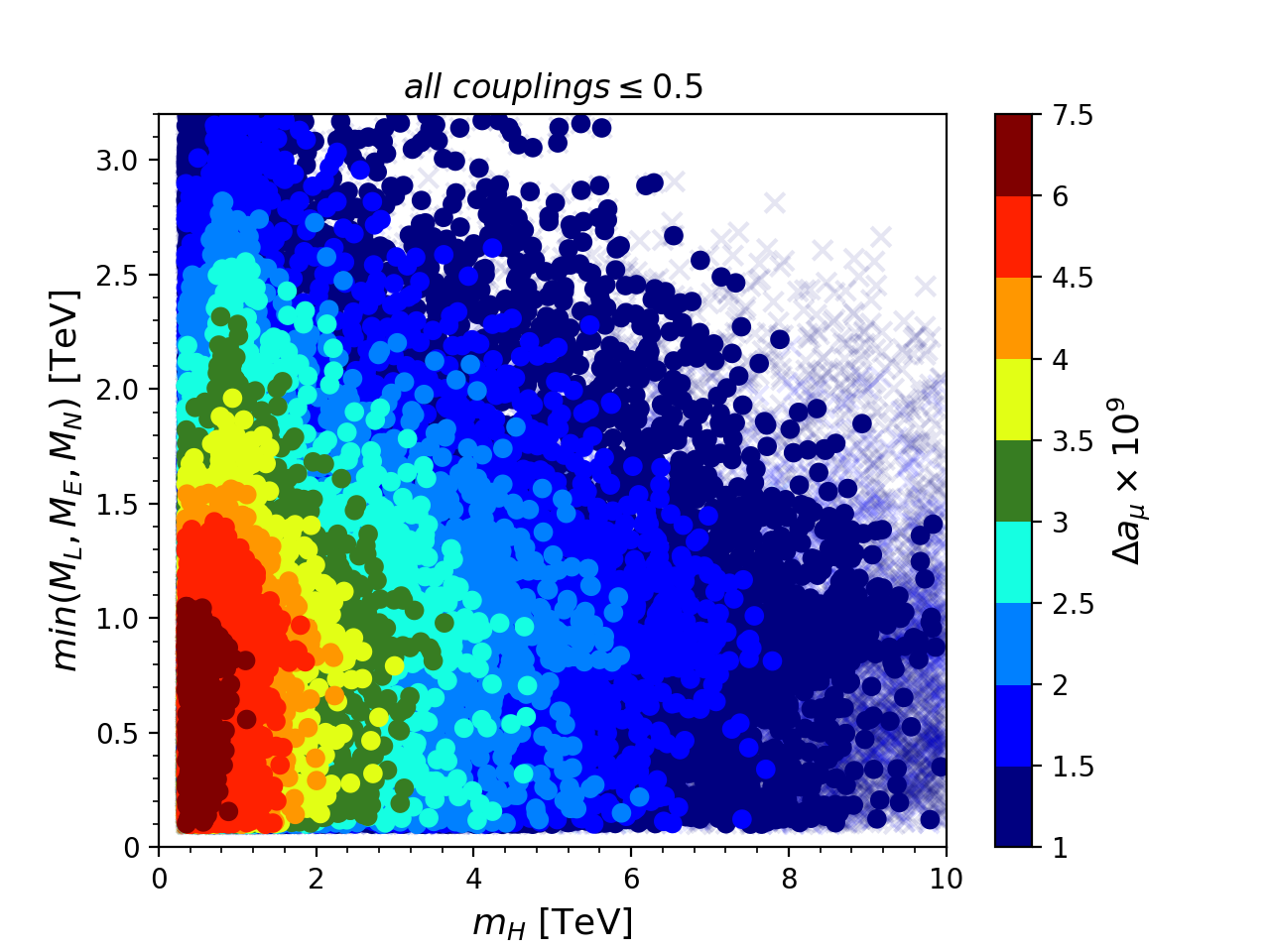}
\includegraphics[scale=0.47]{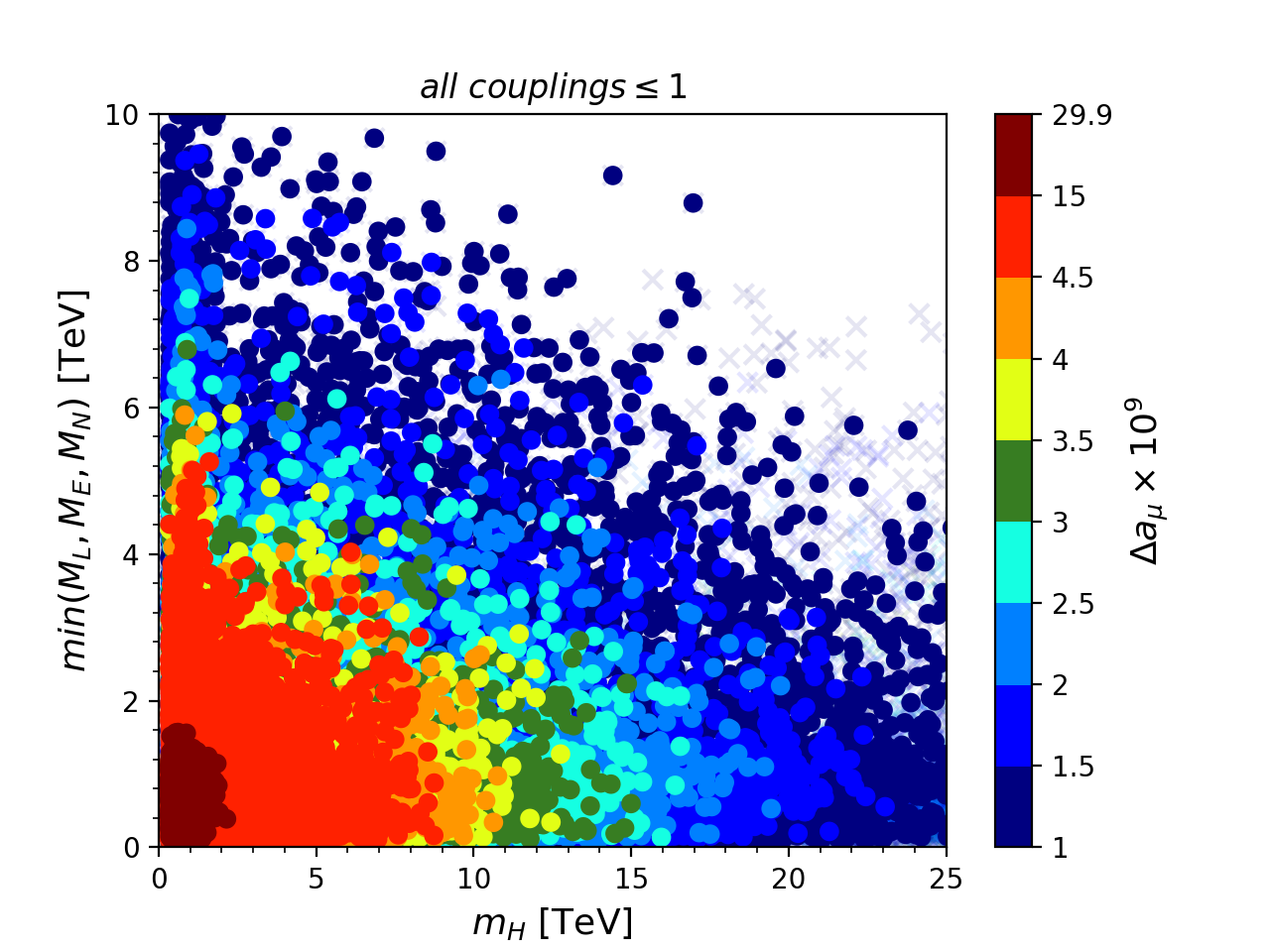}
\caption{Contours of $\Delta a_{\mu}$ in the 2HDM-$Z_{2}$ with respect to $m_{H}$ and $min(M_{L},M_{E},M_{N})$. All values of couplings allowed by precision EW constraints are scanned up to 0.5 (left) and up to 1(right). Lightly shaded crosses without filled circles correspond to scenarios where contributions from heavy Higgses make up less than $50\%$ of the total contribution. Points with larger values of $\Delta a_{\mu}$ are plotted on top.}
\label{fig:gm2_plane}
\end{figure}
\begin{figure}[h]
\centering
\includegraphics[scale=0.23]{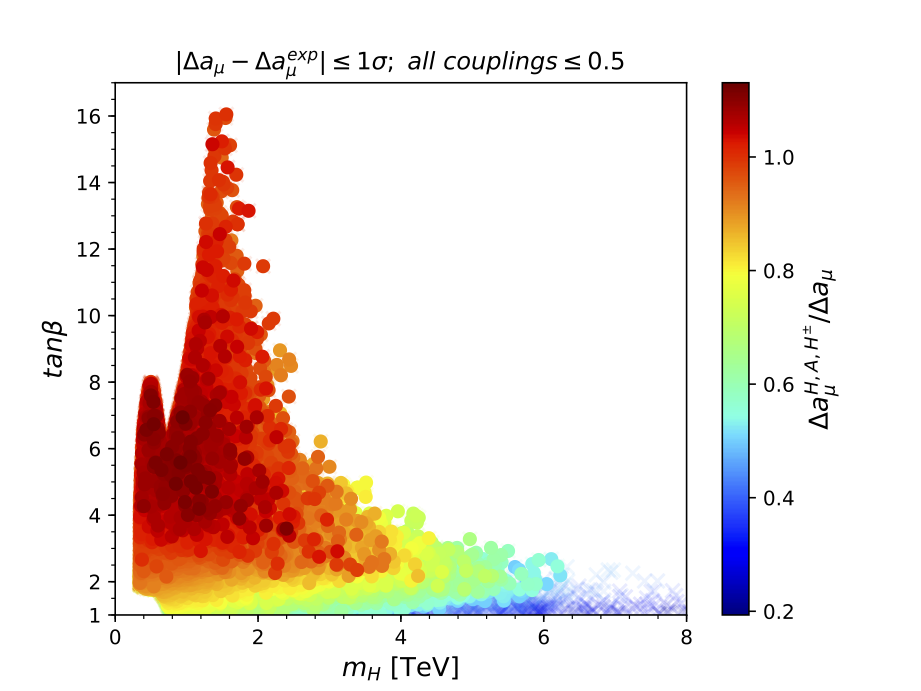}
\includegraphics[scale=0.23]{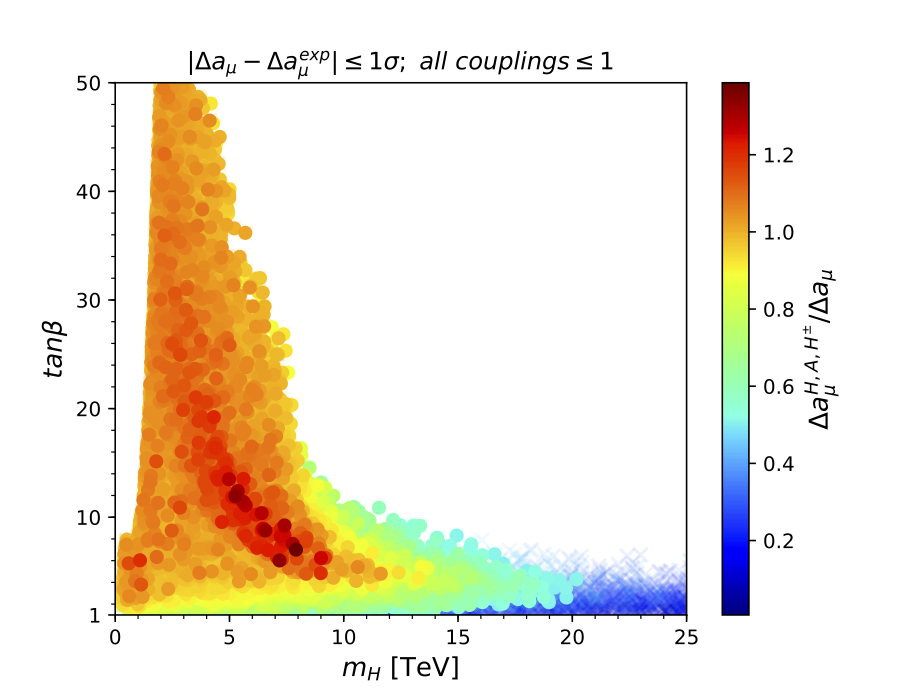}
\caption{Contours of the relative contribution to $\Delta a_{\mu}$ from heavy Higgses, $\Delta a_{\mu}^{H,A,H^{\pm}}/\Delta a_{\mu}$, for values of couplings up to 0.5 (left) and up to 1 (right) when $\Delta a_{\mu}^{exp}$ is achieved within 1$\sigma$ and all constraints are satisfied. Lightly shaded crosses without filled circles correspond to scenarios where contributions from heavy Higgses make up less than $50\%$ of the total contribution. Points with larger values of $\Delta a_{\mu}^{H,A,H^{\pm}}/\Delta a_{\mu}$ are plotted on top.}
\label{fig:dela_mh_tanb}
\end{figure}
In the left panel of Fig.~\ref{fig:gm2_plane}, we show the predicted $\Delta a_{\mu}\times10^{9}$ in the range $1-7.5$ for couplings up to 0.5 with respect to $m_{H}$ and min$(M_{L},M_{E},M_{N})$. In the right panel we show similar contours when the upper bound on couplings is extended to one and predicted values of $\Delta a_{\mu}\times10^{9}$ within $1-29.9$. In both panels, $\tan\beta$ is scanned within $1-50$ assuming constraints on $m_{H}$ from direct searches. For couplings up to 0.5, Higgs masses up to 6 TeV and the lightest new lepton mass up to 3 TeV are viable to explain $\Delta a_{\mu}^{exp}$ within about 1$\sigma$. Assuming all values of couplings not exceeding 1 these ranges extend to 20 TeV and about 8.5 TeV, respectively. Note that the new lepton masses extend to slightly larger values than without up-type couplings, see Ref.~\cite{Dermisek:2020cod}.
\begin{figure}[t]
\centering
\includegraphics[scale=0.23]{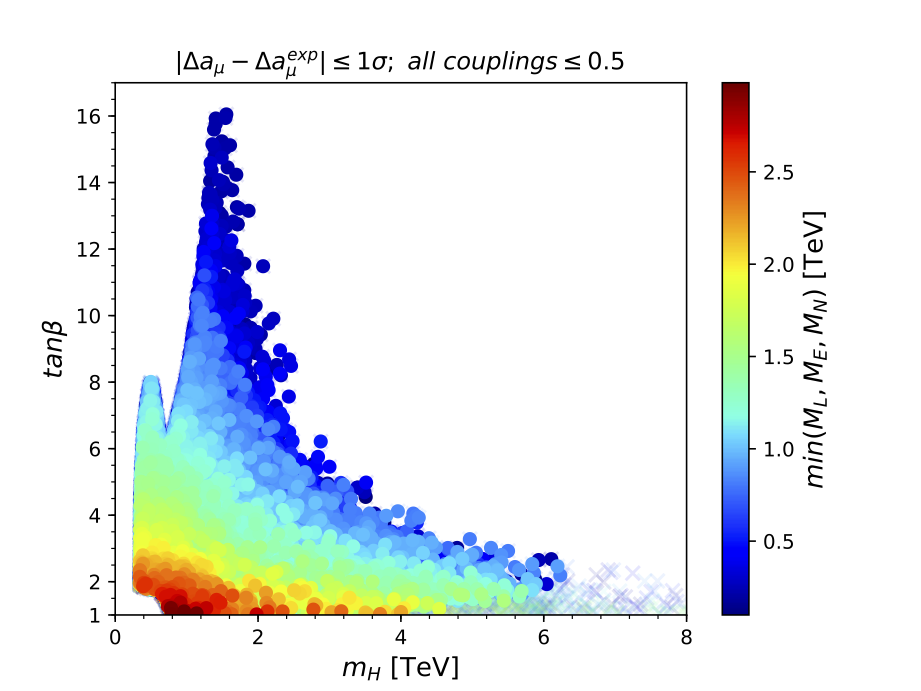}
\includegraphics[scale=0.23]{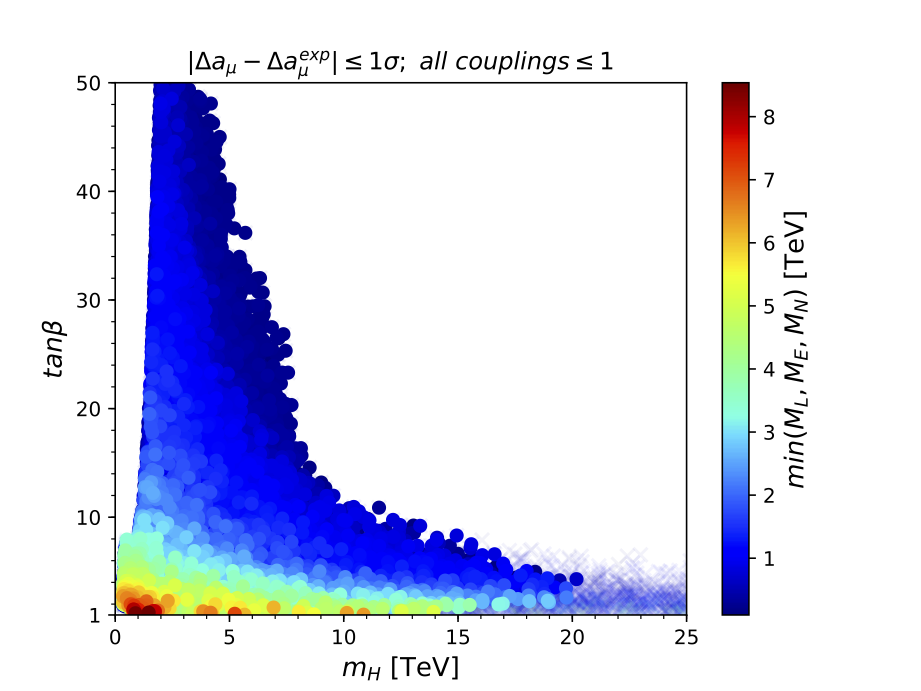}
\caption{Values of $min(M_{L},M_{E},M_{N})$ with respect to $m_{H}$ and $\tan\beta$ for scenarios where $\Delta a_{\mu}^{exp}$ is achieved within 1$\sigma$ for couplings up to 0.5 (left) and 1 (right). Lightly shaded crosses without filled circles correspond to scenarios where contributions from heavy Higgses make up less than $50\%$ of the total contribution. Points with larger values of $min(M_{L},M_{E},M_{N})$ are plotted on top.}
\label{fig:dela_mH_ML}
\end{figure}
In previous sections we highlighted the fact that contributions from heavy Higgs bosons can dominate the total correction to the magnetic moment in most of the parameter space largely due to the $\tan^{2}\beta$ enhancement. In Fig.~\ref{fig:dela_mh_tanb}, we show contributions to $\Delta a_{\mu}$ for couplings up to 0.5 (left) and 1 (right) from heavy Higgses relative to the total contribution with respect to $m_{H}$ and $\tan\beta$ when $\Delta a_{\mu}^{exp}$ is achieved within 1$\sigma$. Lightly shaded crosses correspond to scenarios where heavy Higgses contribute less than $50\%$ to the total correction. We see that the heavy Higgs corrections are generically the largest with increasing $\tan\beta$ and fall off with increasing $m_{H}$. However, we see that for couplings up to 0.5 heavy Higgs masses even up to 6 TeV can give the dominant contribution, while for couplings not exceeding 1 this extends to Higgs masses up to about 20 TeV. We note that regions where $\Delta a_{\mu}^{H,A,H^{\pm}}/\Delta a_{\mu}>1$ occur when the combined contribution from $Z$, $W$, and $h$ is negative.

 In Fig.~\ref{fig:dela_mH_ML}, we show the corresponding range of masses in the same plane as Fig.~\ref{fig:dela_mh_tanb}. Here the range of viable vectorlike lepton masses to at least 3 (8.5) TeV assuming couplings not exceeding 0.5 (1) are explicit. While these upper ranges may be out of reach for future colliders, similar comments as made in~\cite{Dermisek:2020cod} also apply here, where complementary information on precision observables can be used indirectly to fully explore the model. We note that the predictions for modifications of $Z$ and $h$ couplings are almost identical in the present case. Modifications of the $W$ coupling are typically smaller than those of $Z$. However, since $W$ can also receive sizable modifications through up-type couplings it can be larger than the modifications to the $Z$ couplings especially in regions of parameters when the charged Higgs gives the dominant contributions to $\Delta a_{\mu}$.

\subsection{2HDM-II-S with vectorlike leptons}
We remarked in section~\ref{sec:model} that the supersymmetric version of the model in the limit of heavy superpartners has similar structure up to $\bar{\lambda}$ and $\bar{\kappa}$ couplings. In Appendix~\ref{app:SUSY_model} we provide the corresponding approximate formulas for individual contributions to $(g-2)_{\mu}$. The heavy Higgs contributions in the 2HDM-II-S contain both $\tan\beta$ enhanced and suppressed pieces as before. However, the $\tan\beta$ enhanced pieces of these contributions tend to cancel in the leading approximation for $M_{L},M_{E},M_{N}\simeq m_{A}$. Further, the total contribution from $Z$, $W$, and $h$ loops tends to cancel that from heavy Higgses in this limit.

\begin{figure}[t]
\centering
\includegraphics[scale=0.475]{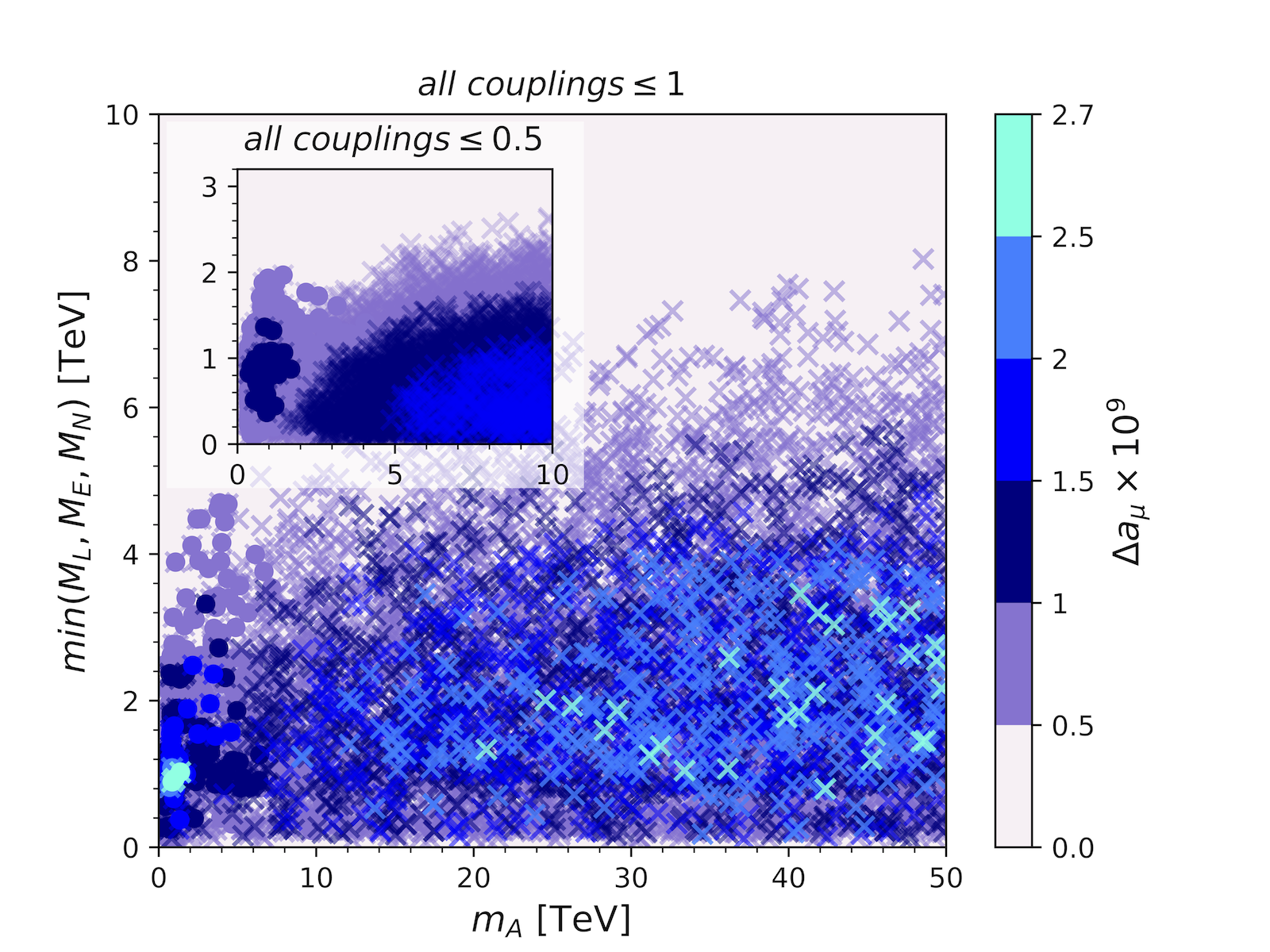}
\includegraphics[scale=0.475]{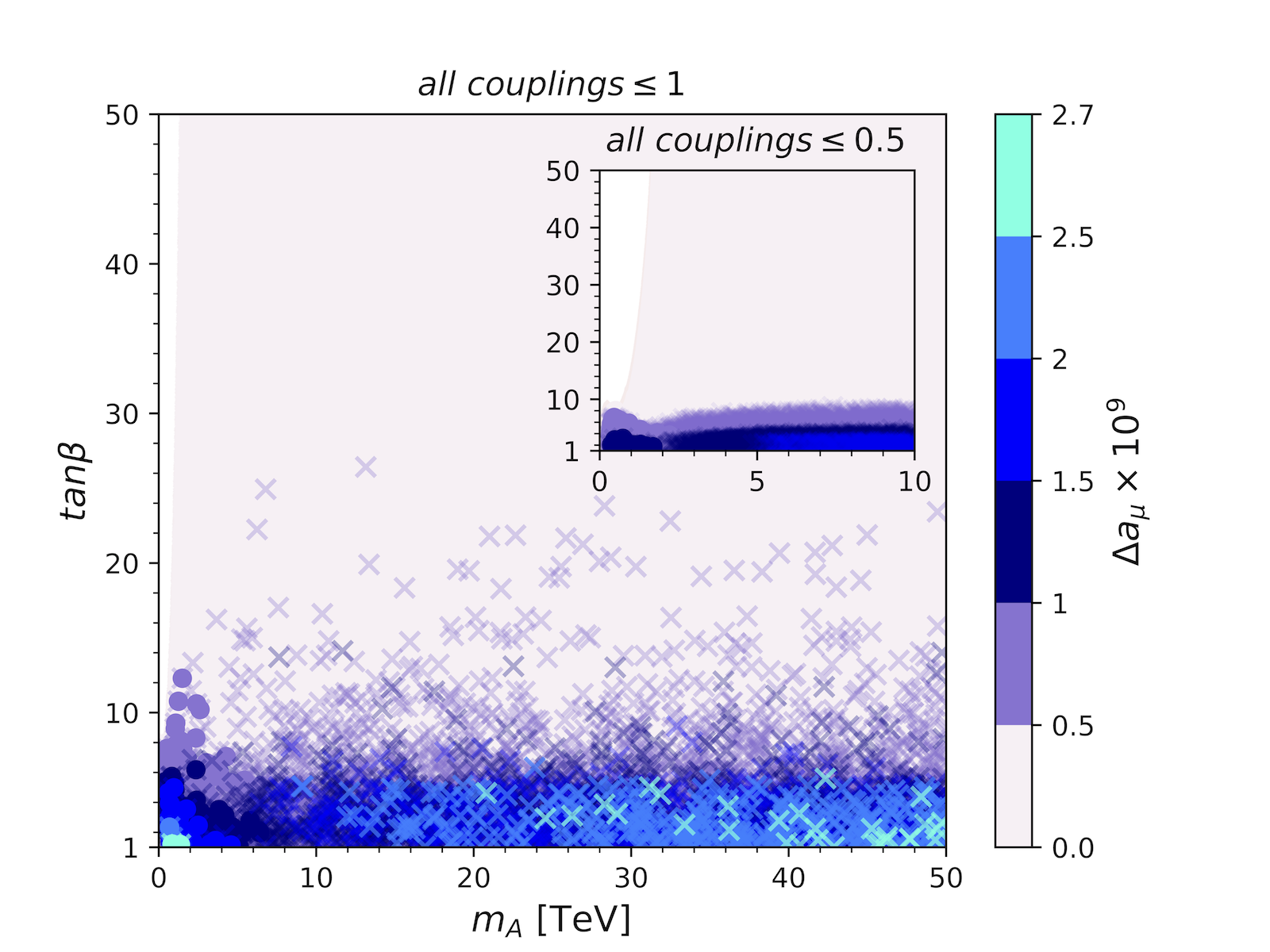}
\caption{{\textbf{Left:}} Predicted values of $\Delta a_{\mu}$ in the 2HDM-II-S with respect to $m_{A}$ and $min(M_{L},M_{E},M_{N})$ for couplings up to 1. The inset shows scenarios when couplings are limited to be less than 0.5. {\textbf{Right:}} The same points with respect to $m_{A}$ and $\tan\beta$. In both panels, shaded crosses without filled circles correspond to scenarios where contributions from heavy Higgses make up less than $50\%$ of the total contribution and points with larger $\Delta a_{\mu}$ are plotted on top.}
\label{fig:susy_gm2}
\end{figure}

In Fig.~\ref{fig:susy_gm2}, we show the total contributions to $\Delta a_{\mu}$ in the 2HDM-II-S with respect to $m_{A}$ and $min(M_{L},M_{E},M_{N})$ (left) and $m_{A}$ and $\tan\beta$ (right) for couplings up to 1. As a result of the cancellation mentioned above the total contribution is smaller than that of the 2HDM-II-$Z_{2}$ over most of the plane. However, it is worth noting that the performance of the model improves as heavy Higgses are decoupled and the total contribution is dominated by $Z$, $W$, and $h$ bosons indicated by the crosses without filled circles. For instance, the model can achieve $\Delta a_{\mu}^{exp}$ within 1$\sigma$ when $m_{A}\gtrsim 9$ TeV and heavy lepton masses lower than about 5 TeV. We note, however, that the discussion of the charged Higgs contribution with respect to up-type couplings also applies in the 2HDM-II-S. Thus, this contribution can dominate in certain regions of parameters. Such scenarios can be seen in the left corner of either panel in Fig.~\ref{fig:susy_gm2} with $m_{A}\lesssim 10$ TeV where contributions from heavy Higgses make up more than 50\% of the total contribution indicated by points with filled circles.
To summarize, the 2HDM-II-S performs less favorably than the 2HDM-II-$Z_{2}$ version with respect to $(g-2)_{\mu}$ considering the loops in Fig.~\ref{fig:diags} largely due to the cancellation of $\tan\beta$ enhanced contributions. However, it should be stressed that the contributions presented in Fig.~\ref{fig:susy_gm2} can be considered \textit{in addition} to the usual contributions from superpartners, e.g. through chargino/sneutrino or neutralino/slepton loops~\cite{Moroi:1995yh,Endo:2011xq,Endo:2011mc,Choudhury:2017fuu}.\\

\subsection{SM with vectorlike leptons}

The standard model with vectorlike leptons was previously studied in~\cite{Kannike:2011ng,Dermisek:2013gta} as an explanation for $\Delta a_{\mu}^{exp}$. Here we extend the region of parameters considered in the model and show the impact of recent measurements of $h\rightarrow \mu^{+}\mu^{-}$. We also explore the correlation of the contribution to $\Delta a_{\mu}$ with modifications of gauge and Yukawa couplings.

\begin{figure}[t]
\centering
\includegraphics[scale=0.23]{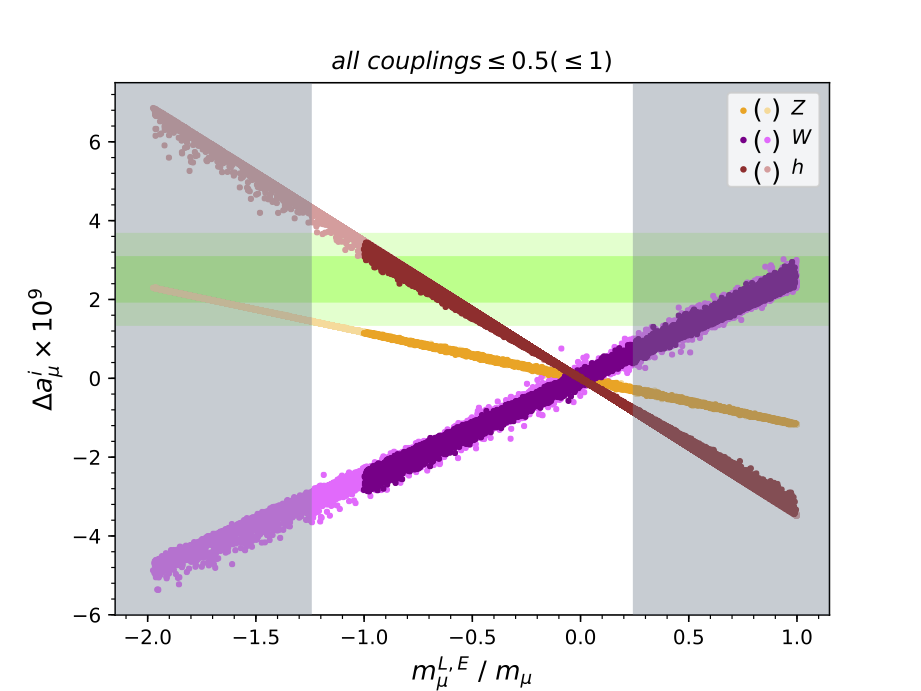}
\includegraphics[scale=0.23]{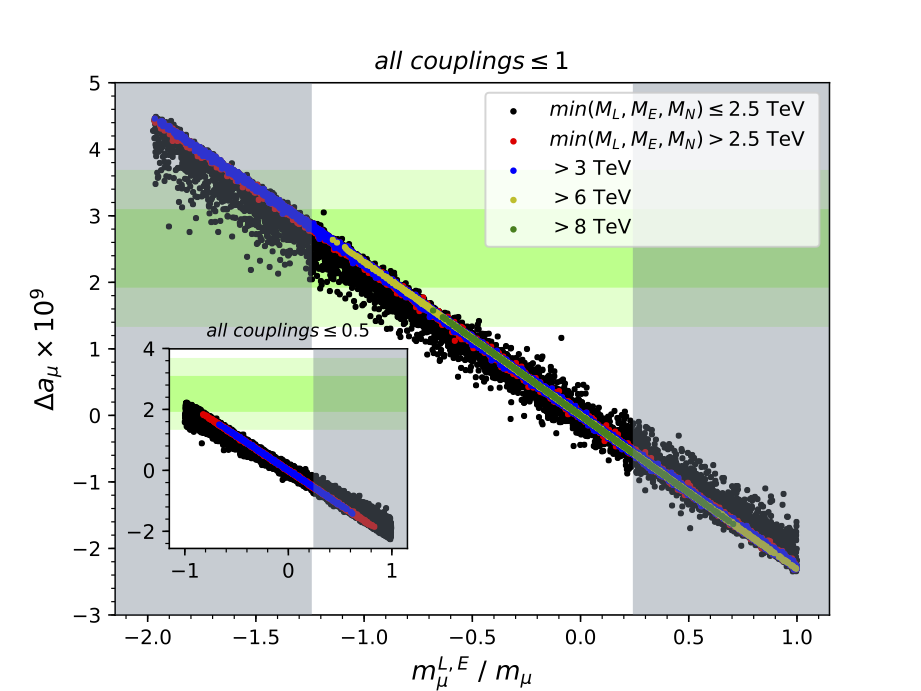}
\caption{{\textbf{Left:}} Individual contributions to $\Delta a_{\mu}$ in the SM with VL's with respect to $m_{\mu}^{LE}/m_{\mu}$.  All values of couplings allowed by constraints up to 1 are shown, where darker shades of corresponding colors show the impact of limiting the upper range of couplings to 0.5. The shaded gray bands show regions of parameters that are ruled out by $h\rightarrow \mu^{+}\mu^{-}$. {\textbf{Right:}} Total contribution to $\Delta a_{\mu}$ from the same scenarios assuming couplings up to 1. Subset of scenarios with couplings up to 0.5 are shown in the inset. We show red, blue, yellow, and green points when $min(M_{L},M_{E},M_{N})>2.5,3,6,$ and 8 TeV respectively and black points when $min(M_{L},M_{E},M_{N})\leq 2.5$ TeV.}
\label{fig:SM_gm2}
\end{figure}

 In Fig.~\ref{fig:SM_gm2}, we show individual contributions to $\Delta a_{\mu}$ with respect to $m_{\mu}^{LE}/m_{\mu}$. All values of couplings allowed by constraints are scanned up to 1 and the dark shades of corresponding colors show the subset of predictions when the upper limit of couplings is reduced to 0.5. In the right panel we show the total contribution of the model to $\Delta a_{\mu}$ with respect to $m_{\mu}^{LE}/m_{\mu}$. Colors represent various lepton masses. The gray shaded bands show the regions of parameters that are ruled out by $h\rightarrow \mu^{+}\mu^{-}$. We see that the SM extended with VL's remains viable as an explanation for $\Delta a_{\mu}^{exp}$ even within 1$\sigma$ for lepton masses as heavy as $\sim7.5$ (2.5) TeV when couplings are allowed up to 1 (0.5). Although we include up-type couplings and $M_{N}$ in the scan, contributions from these parameters have a negligible impact on the results.

\begin{figure}[t]
\centering
\includegraphics[scale=0.23]{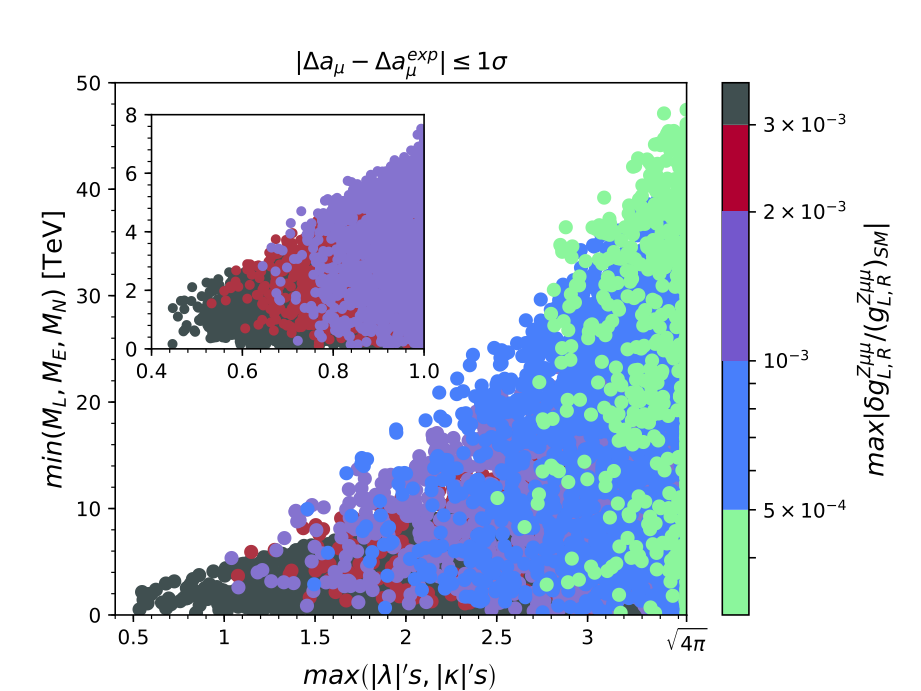}
\includegraphics[scale=0.23]{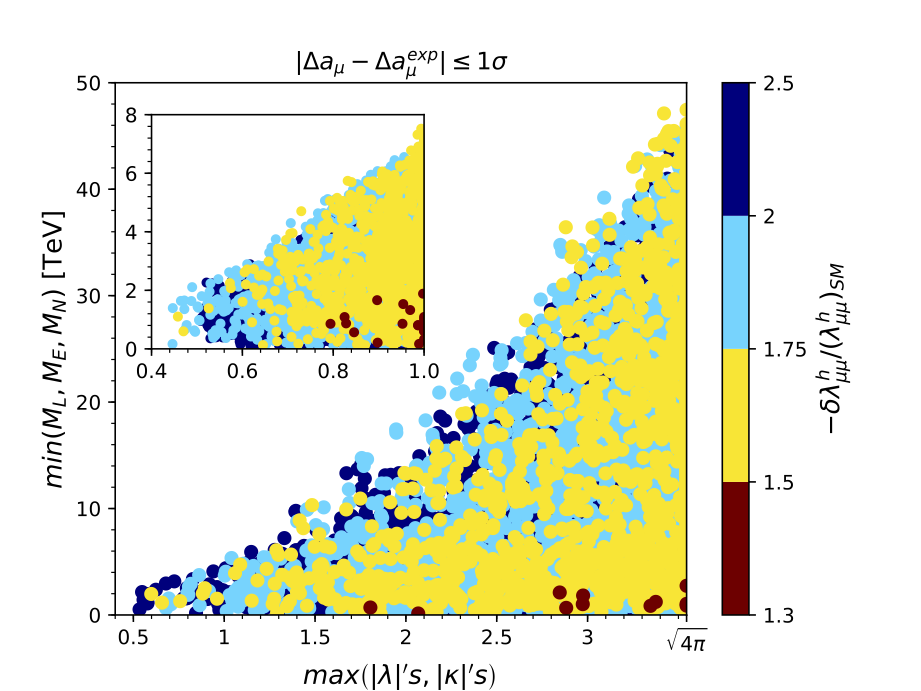}
\caption{\textbf{Left:} Largest relative deviations of the muon couplings to the $Z$-boson in the SM+VL for scenarios leading to an explanation of $\Delta a_{\mu}^{exp}$ within 1$\sigma$. \textbf{Right:} Same scenarios as in the left panel where points show the relative deviation of the Higgs coupling to the muon. In both panels, points leading to the smallest possible deviation from SM couplings are plotted on top.}
\label{fig:SM_couplings}
\end{figure}

Extending the couplings to $\sqrt{4\pi}$, in Fig.~\ref{fig:SM_couplings}, we see that the lightest new lepton mass can be up to 48 TeV while explaining $\Delta a_{\mu}^{exp}$ within 1$\sigma$. Despite the fact that the upper range of masses leading to an explanation of $\Delta a_{\mu}^{exp}$ may be out of reach from direct searches at the LHC, the model can be indirectly probed at future colliders through precision measurements of SM couplings. In colors, we show the deviation of the $Z$-boson couplings to the muon (left), and that of the SM Higgs (right). The insets focus on scenarios when the upper range of couplings is limited to one. Interestingly, the full range of scenarios in the SM with VL's that can explain $\Delta a_{\mu}^{exp}$ within 1$\sigma$ can be indirectly probed by precision measurements at future machines. In particular, the 250 GeV ILC can probe the $Z$-boson couplings up to $\sim5\times 10^{-4}$ which covers almost the entire plane, while the GigaZ option, with sensitivity up to $10^{-4}$, can probe all scenarios up to the perturbativity limit ($\sqrt{4\pi}$).
We note that while the contribution to $\Delta a_{\mu}$ in the 2HDM-II-$Z_{2}$ can be significantly larger than that in the SM with VL's (see for instance an order of magnitude difference between Fig. (\ref{fig:scan_plot}) and Fig. (\ref{fig:SM_gm2}) assuming couplings up to one), the reach of heavy lepton masses able to explain the anomaly is similar, about 45 TeV for couplings up to  $\sqrt{4\pi}$ (compare for example Fig. (\ref{fig:SM_couplings})  with the results in ~\cite{Dermisek:2020cod}). This can be understood by the $\tan\beta$ dependence in the impact of precision EW constraints on the 2HDM-II-$Z_{2}$. For instance, when $\lambda_{E}$ and $\lambda_{L}$ are given by their maximum allowed values, we have $\lambda_{E}=0.03M_{E}/v_{d}$ and $\lambda_{L}=0.04M_{L}/v_{d}$ resulting in $\sin\beta\tan\beta$ enhancement in the heavy Higgs contributions so long as $\lambda_{L,E}$ are smaller than the chosen upper limit. This can be seen in the right panel of Fig.~\ref{fig:line_plot} (where $\lambda_{L,E}<0.5$ is implemented). However, for fixed values of the masses, at some value of $\tan\beta$ the maximum allowed values of couplings are the same as the chosen upper limit (seen from the kinks in Fig.~\ref{fig:line_plot})  and for any larger $\tan\beta$ the constraints have no impact. Thus, the $\sin\beta\tan\beta$ enhancement occurs for lower values of VL masses and moderate values of $\tan\beta$ (up to the kink). However, in order to  explain the measured value of $\Delta a_{\mu}$, for chosen upper limit of couplings the largest possible masses are such that the constraints from precision EW data are automatically satisfied for any $\tan\beta= 1-50$, see Fig. (4) of~\cite{Dermisek:2020cod}. 
\section{Conclusions}
\label{sec:conclusions}
The anomalous measurement of the magnetic moment of the muon remains one of the longest standing deviations of SM predictions. The recent confirmation of this result by the Fermilab Muon g-2 experiment~\cite{Abi:2021gix}, further motivates the presence of physics beyond the SM in nature. In this paper, we studied three distinct extensions of the SM which can provide an explanation of the magnetic moment with heavy new leptons while satisfying low-energy precision measurements.

We extensively explored the viable parameter space of the 2HDM-II-$Z_{2}$, highlighting the range of heavy Higgs and lepton masses which lead to a prediction of $\Delta a_{\mu}$ within 1$\sigma$ of the measured value with model couplings not exceeding 1. This extends the study presented in~\cite{Dermisek:2020cod} for a 2HDM with charged vectorlike leptons by including vectorlike lepton neutral singlets and couplings to $H_{u}$. In particular, we find that the range of lightest lepton and Higgs masses which can explain $\Delta a_{\mu}^{exp}$ within 1$\sigma$ can be as large as 3 (8.5) TeV and 6 (20) TeV, where the dominant contributions are given by loops with heavy Higgses and VL and assuming couplings up to 0.5 (1). 

Notably, these ranges are similar to the study without up-type couplings~\cite{Dermisek:2020cod}. However, if the typically dominant contributions from down-type couplings are not present the contribution from the charged Higgs itself can still explain $\Delta a_{\mu}^{exp}$ within 1$\sigma$ due to the presence of couplings to $H_{u}$. Apart from the main results, we also emphasize that the model can generate $\Delta a_{\mu}$ one (two) orders of magnitude larger than the central measured value with couplings up to 1 ($\sqrt{4\pi}$) while satisfying all current precision constraints. While it is expected that even the LHC running at 14 TeV with $3ab^{-1}$ luminosity can only exclude (doublet) VL masses up to 1250 GeV~\cite{Bhattiprolu:2019vdu} (depending on the decay modes), the high range of masses we present here can be probed indirectly at future precision machines~\cite{Dermisek:2020cod}. 

In addition to our study of the 2HDM-II-$Z_{2}$, we emphasize that while Yukawa couplings of SM leptons in this model are indistinguishable to those in the MSSM, couplings of VL are necessarily different due to the requirement that the superpotential be holomorphic. This leads to drastically different results in the contributions to $\Delta a_{\mu}$ from the same particle content. We find that the 2HDM-II-S can typically generate $\Delta a_{\mu}$ within 1$\sigma$ from the central measured value in the limit that heavy Higgses are decoupled, for $m_{A}\gtrsim 9$ TeV, as a result of cancellations between heavy Higgs contributions and those of $Z$, $W$, and $h$. However, in a subset of the parameter space the charged Higgs contribution can even reach the central value in the presence of sizable (up to 1) couplings to $H_{u}$. 

For completeness we have extended previous studies of the SM with VL~\cite{Kannike:2011ng,Dermisek:2013gta} by including couplings to heavy leptons which are SM singlets. Interestingly, we find that while the reach of lepton masses which can lead to $\Delta a_{\mu}$ within 1$\sigma$ from the central measured value is roughly the same as in the 2HDM-II-$Z_{2}$, current measurements of $h\rightarrow \mu^{+}\mu^{-}$ have a much bigger impact in this scenario limiting the possible contribution to $\Delta a_{\mu}$ up to the current central value. This can be understood from the $\tan\beta$ dependence in the impact of precision EW constraints on the 2HDM-II-$Z_{2}$. This impact also appears in the differences in modifications of the $Z$ and $h$ couplings to the muon required to explain $\Delta a_{\mu}^{exp}$ within 1$\sigma$. For the SM with vectorlike leptons we find that the 250 GeV ILC, that can probe the $Z$-boson couplings up to $\sim5\times 10^{-4}$, can cover almost all the parameter space, while the GigaZ option, with sensitivity up to $10^{-4}$, can probe all scenarios up to the perturbativity limit ($\sqrt{4\pi}$). This is in sharp contrast to the predictions of the 2HDM-II-$Z_{2}$~\cite{Dermisek:2020cod}.

\acknowledgments
We thank Nima Arkani-Hamed and Keisuke Harigaya for useful discussions. The work of RD was supported in part by the U.S. Department of Energy under grant number {DE}-SC0010120. NM acknowledges partial support by the U.S. Department of Energy under contracts No. DEAC02-06CH11357 at
Argonne National Laboratory. TRIUMF receives federal funding via a contribution agreement with the National Research Council of Canada.

\appendix
\section{Couplings and approximate formulas in the 2HDM-II-$Z_{2}$}
\label{sec:Z2_appendix}

We consider a complete generation of VL's which can mix with the 2nd generation leptons of the SM. In section \ref{sec:loops}, we present one-loop formulas giving contributions to $(g-2)_{\mu}$ in a generic 2HDM. In the following appendices we derive general expressions for all relevant couplings in the 2HDM-II-$Z_{2}$ we consider and present useful approximations for individual couplings in the limit of heavy lepton masses.

\subsection{Couplings to $Z$ and $W$ bosons}
Expressions for couplings of charged and neutral leptons to $Z$ and $W$ bosons have been given previously in the SM extended with vectorlike leptons, and in the 2HDM-II-$Z_{2}$~\cite{Dermisek:2013gta, Dermisek:2015oja}. We summarize these expressions for completeness.\footnote{Note that since the $U(1)_{EM}$ charges of the vectorlike leptons are the same as their SM counterparts, couplings to the photon are not modified by mixing.} In the following it will be convenient to define the 3-component vectors $e_{L,Ra}\equiv(\mu_{L,R}, L_{L,R}^{-}, E_{L,R})^{T}$, and $\nu_{L,Ra}\equiv((\nu_{\mu})_{L,R}, L_{L,R}^{0}, N_{L,R})^{T}$ in the gauge eigenstate basis. We denote 3-vectors of mass eigenstates by $e_{L,R}=U^{e}_{L,R}\hat{e}_{L,R}$ and $\nu_{L,R}=U^{\nu}_{L,R}\hat{\nu}_{L,R}$, where $U^{e}_{L,R}$ and $U^{\nu}_{L,R}$ are the diagonalization matrices given by Eqs. (\ref{eq:mixing_matrices_1}) and (\ref{eq:mixing_matrices_2}). We label the components of mass eigenstate vectors by $a=2,4,5$.

The couplings to the $Z$ bosons follow from the kinetic terms of leptons:
\begin{flalign}
\label{eq:kin_lag}\nonumber
\mathcal{L}_{kin}\supset \;& \bar{e}_{La}i\slashed{D}_{a}e_{La} + \bar{e}_{Ra}i\slashed{D}_{a}e_{Ra} + \bar{\nu}_{La}i\slashed{D}_{a}\nu_{La} + \bar{\nu}_{Ra}i\slashed{D}_{a}\nu_{Ra}\\\nonumber
 =\;&  \bar{\hat{e}}_{La}(U_{L}^{e\dagger})_{ac}i\slashed{D}_{c}(U_{L}^{e})_{cb}\hat{e}_{Lb} + \bar{\hat{e}}_{Ra}(U_{R}^{e\dagger})_{ac}i\slashed{D}_{c}(U_{R}^{e})_{cb}\hat{e}_{Rb}\\
	& + \bar{\hat{\nu}}_{La}(U_{L}^{\nu\dagger})_{ac}i\slashed{D}_{c}(U_{L}^{\nu})_{cb}\hat{\nu}_{Lb} + \bar{\hat{\nu}}_{Ra}(U_{R}^{\nu\dagger})_{ac}i\slashed{D}_{c}(U_{R}^{\nu})_{cb}\hat{\nu}_{Rb},
\end{flalign}
where the covariant derivative is given by
\begin{equation}
\label{eq:cov_deriv}
D_{\mu a}=\partial_{\mu} - i\frac{g}{\cos\theta_{W}}(T^{3}_{a} - \sin^{2}\theta_{W}Q_{a})Z_{\mu}.
\end{equation}
Defining the couplings of the $Z$ boson to leptons $f_a$ and $f_b$ as
\begin{equation}
\label{eq:Z_lag}
\mathcal{L}\supset (\bar{f}_{La}\gamma^{\mu}g_{L}^{Zf_{a}f_{b}}f_{Lb} + \bar{f}_{Ra}\gamma^{\mu}g_{R}^{Zf_{a}f_{b}}f_{Rb})Z_{\mu},
\end{equation}
the couplings of left- and right- handed fields immediately follow from Eq. \ref{eq:cov_deriv}
\begin{flalign}
\label{eq:Z_couplings}
g_{L}^{Ze_{a}e_{b}}& = \frac{g}{\cos\theta_W}\left[\left(-\frac{1}{2}+\sin^2\theta_{W}\right)\delta_{ab} + \frac{1}{2}(U_{L}^{e\dagger})_{a5}(U_{L}^{e})_{5b}\right],\\
g_{R}^{Ze_{a}e_{b}} & =  \frac{g}{\cos\theta_W}\left[\sin^2\theta_{W}\delta_{ab} - \frac{1}{2}(U_{R}^{e\dagger})_{a4}(U_{R}^{e})_{4b}\right],\\
g_{L}^{Z\nu_{a}\nu_{b}} & = \frac{g}{2\cos\theta_W}\left[\delta_{ab} - (U_{L}^{\nu\dagger})_{a5}(U_{L}^{\nu})_{5b}\right],\\
g_{R}^{Z\nu_{a}\nu_{b}} &= \frac{g}{2\cos\theta_{W}}(U_{R}^{\nu\dagger})_{a4}(U_{R}^{\nu})_{4b},
\end{flalign}
where $a,b=2,4,5$. Since we only introduce mixing to the muon and muon neutrino, couplings of the first and third generation leptons in the SM are not modified.

The couplings of the $W$ boson to charged and neutral leptons arise from the kinetic terms
\begin{flalign}
\label{eq:W_kin}
\mathcal{L}_{kin} \supset \;&\frac{g}{\sqrt{2}}(\bar{\nu}_{\mu}\gamma^{\mu}\mu_{L} + \bar{L}^{0}_{L}\gamma^{\mu}L^{-}_{L} + \bar{L}^{0}_{R}\gamma^{\mu}L^{-}_{R})W^{+}_{\mu} + h.c\\\nonumber
=\;& \frac{g}{\sqrt{2}}\Bigg[\bar{\hat{\nu}}_{La}(U_{L}^{\nu\dagger})_{a2}\gamma^{\mu}(U_{L}^{e})_{2b}\hat{e}_{Lb} + \bar{\hat{\nu}}_{La}(U_{L}^{\nu\dagger})_{a4}\gamma^{\mu}(U_{L}^{e})_{4b}\hat{e}_{Lb}\\
& \hspace{1cm}+ \bar{\hat{\nu}}_{Ra}(U_{R}^{\nu\dagger})_{a4}\gamma^{\mu}(U_{R}^{e})_{4b}\hat{e}_{Lb} \Bigg]W^{+}_{\mu} + h.c.
\end{flalign}
Defining the couplings of the $W$ boson to mass eigenstates $\hat{\nu}_a$ and $\hat{e}_a$ as
\begin{equation}
\label{eq:W_lag}
\mathcal{L}\supset \left(\bar{\hat{\nu}}_{La}\gamma^{\mu}g_{L}^{W\nu_{a}e_{b}}\hat{e}_{Lb}+ \bar{\hat{\nu}}_{Ra}\gamma^{\mu}g_{R}^{W\nu_{a}e_{b}}\hat{e}_{Rb}\right)W^{+}_{\mu} + h.c.,
\end{equation}
we find
\begin{flalign}
\label{eq:W_couplings}
g_{L}^{W\nu_{a}e_{b}} & =\frac{g}{\sqrt{2}}\left[(U_{L}^{\nu\dagger})_{a2}(U_{L}^{e})_{2b} + (U_{L}^{\nu\dagger})_{a4}(U_{L}^{e})_{4b}\right],\\
g_{R}^{W\nu_{a}e_{b}} & = \frac{g}{\sqrt{2}} (U_{R}^{\nu\dagger})_{a4}(U_{R}^{e})_{4b}.
\end{flalign}
\subsection{Couplings to Higgs bosons}
Here we provide our conventions for the Higgs sector and couplings of VL leptons to physical Higgs and Goldstone bosons.

In the basis where the Yukawa couplings of SM leptons are diagonal, the Yukawa couplings of the neutral Higgs components to the muon and VL leptons are given by
\begin{flalign}
\label{eq:H_lag}\nonumber
\mathcal{L}_{H^{0}_{u,d}} = - &\left(\bar{\mu}_{L}, \bar{L}_{L}^{-}, \bar{E}_{L}\right)\begin{pmatrix} y_{\mu}H^{0}_{d} & 0 & \lambda_{E}H^{0}_{d}\\ 
						\lambda_{L}H^{0}_{d} & 0 & \lambda H^{0}_{d}\\
						0 & \bar{\lambda}H^{0\dagger}_{d}& 0\end{pmatrix}\begin{pmatrix}\mu_{R}\\ L_{R}^{-}\\ E_{R}\end{pmatrix}\\
		&-(\bar{\nu}_{\mu}, \bar{L}_{L}^{0}, \bar{N}_{L})\begin{pmatrix}0&0 &\kappa_{N}H^{0}_{u}\\0&0&\kappa H^{0}_u\\0&\bar{\kappa}H^{0\dagger}_{u}& 0\end{pmatrix}\begin{pmatrix}\nu_{R}=0\\ L_{R}^{0}\\ N_{R}\end{pmatrix}.				
\end{flalign}
To write these interactions in terms of mass eigenstates, we additionally rotate the Higgs fields to the basis where physical and Goldstone degrees of freedom are apparent. This basis is defined by
\begin{flalign}
H^{0}_{d} =& v_{d} + \frac{1}{\sqrt{2}}(-h\sin\alpha + H\cos\alpha) +\frac{i}{\sqrt{2}}(G\cos\beta - A\sin\beta),\\
H^{0}_{u} =& v_{u} + \frac{1}{\sqrt{2}}(h\cos\alpha + H\sin\alpha) -\frac{i}{\sqrt{2}}(G\sin\beta + A\cos\beta),
\end{flalign}
and for the charged sector
\begin{flalign}
H_{d}^{\pm}=&\cos\beta G^{\pm} - \sin\beta H^{\pm},\\
H_{u}^{\pm} =& -\sin\beta G^{\pm} - \cos\beta H^{\pm}.
\end{flalign}
Inverting these relations, the mass eigenstates of the neutral Higgs and Goldstone bosons are given by
\begin{flalign}
\label{eq:nH_states}
\begin{pmatrix}H\\h\end{pmatrix}=&\begin{pmatrix}\cos\alpha & \sin\alpha\\
									   -\sin\alpha & \cos\alpha
							\end{pmatrix}\begin{pmatrix}\sqrt{2}(\text{Re}H_{d}^{0} - v_{d})\\
													\sqrt{2}(\text{Re}H_{u}^{0} - v_{u})
													\end{pmatrix},\\
\begin{pmatrix}G\\A\end{pmatrix}=&\begin{pmatrix}\cos\beta & \sin\beta\\
									   -\sin\beta & \cos\beta
							\end{pmatrix}\begin{pmatrix}\sqrt{2}(\text{Im}H_{d}^{0})\\
													-\sqrt{2}(\text{Im}H_{u}^{0})
													\end{pmatrix}.
\end{flalign}
where $h$ and $H$, $A$, and $G$ are the CP-even, CP-odd, and neutral Goldstone bosons, respectively. By requiring a light Higgs with couplings to gauge bosons that are identical to those in the SM we have $\alpha=\beta-\pi/2$, and the mass eigenstates for $h$ and $H$ are 
\begin{equation}
\label{eq:phys_nH_states}
\begin{pmatrix}h\\-H\end{pmatrix}=\begin{pmatrix}\cos\beta & \sin\beta\\
									   -\sin\beta & \cos\beta
							\end{pmatrix}\begin{pmatrix}\sqrt{2}(\text{Re}H_{d}^{0} - v_{d})\\
													\sqrt{2}(\text{Re}H_{u}^{0} - v_{u})
													\end{pmatrix}.
\end{equation}

Thus, in term of mass eigenstates the Yukawa couplings of charged and neutral leptons to $h$ and $H$ are 
\begin{flalign}
\label{eq:phys_nH_lag}
\mathcal{L}_{h,H} =& - \frac{1}{\sqrt{2}}\bar{\hat{e}}_{L}U_{L}^{e\dagger}Y_{E}U_{R}^{e}\hat{e}_{R}(h\cos\beta + H\sin\beta)\\
			&- \frac{1}{\sqrt{2}}\bar{\hat{\nu}}_{L}U_{L}^{\nu\dagger}Y_{N}U_{R}^{\nu}\hat{\nu}_{R}(h\sin\beta - H\cos\beta) + h.c.,
\end{flalign}
where $Y_{E}$ and $Y_{N}$ are given by
\begin{flalign}
\label{eq:H_Yukawa}
Y_{E}=&\begin{pmatrix}y_{\mu} & 0 & \lambda_{E}\\
					\lambda_{L} & 0 & \lambda\\
					0 & \bar{\lambda} & 0\end{pmatrix},\;\text{and}\;
Y_{N}=\begin{pmatrix}0 & 0 & \kappa_{N}\\
					0 & 0 & \kappa\\
					0 & \bar{\kappa} & 0\end{pmatrix}.		
\end{flalign}
The lagrangian for Yukawa couplings to CP-even Higgses can be written as
\begin{flalign}\nonumber
\mathcal{L}_{h,H} =& - \frac{1}{\sqrt{2}}\bar{\hat{e}}_{La}\lambda^{h}_{e_{a}e_{b}}\hat{e}_{Rb}h - \frac{1}{\sqrt{2}}\bar{\hat{\nu}}_{La}\lambda^{h}_{\nu_{a}\nu_{b}}\hat{\nu}_{Rb}h\\
	& - \frac{1}{\sqrt{2}}\bar{\hat{e}}_{La}\lambda^{H}_{e_{a}e_{b}}\hat{e}_{Rb}H - \frac{1}{\sqrt{2}}\bar{\hat{\nu}}_{La}\lambda^{H}_{\nu_{a}\nu_{b}}\hat{\nu}_{Rb}H + h.c.,
\end{flalign}
where
\begin{flalign}
\label{eq:H_couplings}
\lambda^{h}_{e_{a}e_{b}}=&\cos\beta(U^{e\dagger}_{L}Y_{E}U^{e}_{R})_{ab},\\
\lambda^{h}_{\nu_{a}\nu_{b}}=&\sin\beta(U^{\nu\dagger}_{L}Y_{N}U^{\nu}_{R})_{ab},\\
\lambda^{H}_{e_{a}e_{b}}=&\sin\beta(U^{e\dagger}_{L}Y_{E}U^{e}_{R})_{ab},\\
\lambda^{H}_{\nu_{a}\nu_{b}}=&-\cos\beta(U^{\nu\dagger}_{L}Y_{N}U^{\nu}_{R})_{ab}.
\end{flalign}
The couplings for the CP-odd Higgs can be derived in a similar way. The langrangian for the Yukawa couplings to $A$ reads
\begin{flalign}
\label{eq:A_lag}\nonumber
\mathcal{L}_{A}=&-\frac{i}{\sqrt{2}}\bar{\hat{e}}_{L}U_{L}^{e\dagger}Y_{E}^{A}U_{R}^{e}\hat{e}_{R}(-A\sin\beta)\\
				&-\frac{i}{\sqrt{2}}\bar{\hat{\nu}}_{L}U_{L}^{\nu\dagger}Y_{N}^{A}U_{R}^{\nu}\hat{\nu}_{R}(-A\cos\beta) + h.c.,
\end{flalign}
\label{eq:A_Yukawa}
where $Y^{A}_{E}$ and $Y^{A}_{N}$ are given by
\begin{flalign}
Y_{E}^{A}=\begin{pmatrix}y_{\mu} & 0 & \lambda_{E}\\
					\lambda_{L} & 0 & \lambda\\
					0 & -\bar{\lambda} & 0\end{pmatrix}, \text{and} \hspace{0.25cm}Y_{N}^{A}=&\begin{pmatrix}0 & 0 & \kappa_{N}\\
					0 & 0 & \kappa\\
					0 & -\bar{\kappa} & 0\end{pmatrix}.
\end{flalign}
Writing the lagrangian as 
\begin{equation}
\mathcal{L}_{A}=-\frac{1}{\sqrt{2}}\bar{\hat{e}}_{La}\lambda^{A}_{e_{a}e_{b}}\hat{e}_{Rb}A - \frac{1}{\sqrt{2}}\bar{\hat{\nu}}_{La}\lambda^{A}_{\nu_{a}\nu_{b}}\hat{\nu}_{Rb}A + h.c.,
\end{equation}
we have
\begin{flalign}
\label{eq:A_couplings}
\lambda^{A}_{e_{a}e_{b}} & = -i\sin\beta(U^{e\dagger}_{L}Y_{E}^{A}U^{e}_{R})_{ab},\\
\lambda^{A}_{\nu_{a}\nu_{b}} & = -i\cos\beta(U^{\nu\dagger}_{L}Y_{N}^{A}U^{\nu}_{R})_{ab}.
\end{flalign}

The couplings for the neutral Goldstone boson, $G$, follow similarly. Defining 
\begin{equation}
\label{eq:G_lag}
\mathcal{L}_{G}=-\frac{1}{\sqrt{2}}\bar{\hat{e}}_{La}\lambda^{G}_{e_{a}e_{b}}\hat{e}_{Rb}G - \frac{1}{\sqrt{2}}\bar{\hat{\nu}}_{La}\lambda^{G}_{\nu_{a}\nu_{b}}\hat{\nu}_{Rb}G + h.c.,
\end{equation}
we get 
\begin{flalign}
\label{eq:G_couplings}
\lambda^{G}_{e_{a}e_{b}} & = i\cos\beta(U^{e\dagger}_{L}Y_{E}^{A}U^{e}_{R})_{ab},\\
\lambda^{G}_{\nu_{a}\nu_{b}} & = -i\sin\beta(U^{\nu\dagger}_{L}Y_{N}^{A}U^{\nu}_{R})_{ab}.
\end{flalign}

For couplings to the charged Higgs bosons we first define $H_{d}^{-}\equiv (H_{d}{^{+})^\dagger}$ and $H_{u}^{+}\equiv (H_{u}^{-})^{\dagger}$. Then, reading off the interactions from the lagrangian we get
\begin{flalign}
\label{eq:cH_lag}\nonumber
\mathcal{L}_{H^{\pm}_{u,d}} = - &\left(\bar{\nu}_{\mu}, \bar{L}_{L}^{0}, \bar{N}_{L}\right)\begin{pmatrix} y_{\mu}H^{+}_{d} & 0 & \lambda_{E}H^{+}_{d}\\ 
						\lambda_{L}H^{+}_{d} & 0 & \lambda H^{+}_{d}\\
						0 & \bar{\kappa}H^{+}_{u}& 0\end{pmatrix}\begin{pmatrix}\mu_{R}\\ L_{R}^{-}\\ E_{R}\end{pmatrix}\\
		&-(\bar{\mu}_{L}, \bar{L}_{L}^{-}, \bar{E}_{L})\begin{pmatrix}0&0 &\kappa_{N}H^{-}_{u}\\0&0&\kappa H^{-}_{u}\\0&\bar{\lambda}H^{-}_{d}& 0\end{pmatrix}\begin{pmatrix}0\\ L_{R}^{0}\\ N_{R}\end{pmatrix}.				
\end{flalign}
The charged Higgs mass eigenstates $H^{\pm}$ and $G^{\pm}$ are related to the gauge eigenstates by
\begin{equation}
\label{eq:cH_states}
	\begin{pmatrix}G^{\pm}\\H^{\pm}\end{pmatrix}=\begin{pmatrix}\cos\beta & \sin\beta\\
												    -\sin\beta & \cos\beta \end{pmatrix}\begin{pmatrix}H_{d}^{\pm} \\ -H_{u}^{\pm}\end{pmatrix}.
\end{equation}
Thus, the Yukawa couplings to charged Higgs bosons, in terms of mass eigenstates, are given by
\begin{flalign}
\mathcal{L}_{H^{\pm}}=-\bar{\hat{\nu}}_{L}U^{\nu\dagger}_{L}Y_{N}^{H^{\pm}}U^{e}_{R}\hat{e}_{R}H^{+} - \bar{\hat{e}}_{L}U^{e\dagger}_{L}Y_{E}^{H^{\pm}}U^{\nu}_{R}\hat{\nu}_{R}H^{-} + h.c.,
\end{flalign}
where
\begin{flalign}
\label{eq:cH_Yukawa}
Y_{N}^{H^{\pm}} =& -\sin\beta\begin{pmatrix} y_{\mu} & 0 & \lambda_{E}\\ 
						\lambda_{L}& 0 & \lambda \\
						0 & \bar{\kappa}/\tan\beta& 0\end{pmatrix},\;\text{and}\;
Y_{E}^{H^{\pm}} = -\cos\beta\begin{pmatrix} 0 & 0 & \kappa_{N}\\ 
							  0& 0 & \kappa \\
						0 & \bar{\lambda}\tan\beta& 0\end{pmatrix}.
\end{flalign}

Finally, writing the lagrangian for charged Higgs Yukawa couplings as 
\begin{equation}
\mathcal{L}_{H^{\pm}} = -\bar{\hat{\nu}}_{La}\lambda^{H^{\pm}}_{\nu_{a}e_{b}}\hat{e}_{Rb}H^{+}  - \bar{\hat{e}}_{La}\lambda^{H^{\pm}}_{e_{a}\nu_{b}}\hat{\nu}_{Rb}H^{-} + h.c.,
\end{equation}
we have
\begin{flalign}
\label{eq:cH_couplings}
\lambda^{H^{\pm}}_{\nu_{a}e_{b}} = &(U^{\nu\dagger}_{L}Y_{N}^{H^{\pm}}U^{e}_{R})_{ab},\\
\lambda^{H^{\pm}}_{e_{a}\nu_{b}} = &(U^{e\dagger}_{L}Y_{E}^{H^{\pm}}U^{\nu}_{R})_{ab}.
\end{flalign}

The couplings for the charged Goldstone bosons, $G^{\pm}$, follow similarly. Defining

\begin{equation}
\label{eq:cG_lag}
\mathcal{L}_{G^{\pm}} = -\bar{\hat{\nu}}_{La}\lambda^{G^{\pm}}_{\nu_{a}e_{b}}\hat{e}_{Rb}G^{+}  - \bar{\hat{e}}_{La}\lambda^{G^{\pm}}_{e_{a}\nu_{b}}\hat{\nu}_{Rb}G^{-} + h.c.,
\end{equation}
we get
\begin{flalign}
\label{eq:cG_coulpings}
\lambda^{G^{\pm}}_{\nu_{a}e_{b}} = &(U^{\nu\dagger}_{L}Y_{N}^{G^{\pm}}U^{e}_{R})_{ab},\\
\lambda^{G^{\pm}}_{e_{a}\nu_{b}} = &(U^{e\dagger}_{L}Y_{E}^{G^{\pm}}U^{\nu}_{R})_{ab},
\end{flalign}
where
\begin{flalign}
Y_{N}^{G^{\pm}} =& \cos\beta\begin{pmatrix} y_{\mu} & 0 & \lambda_{E}\\ 
						\lambda_{L}& 0 & \lambda \\
						0 & -\bar{\kappa}\tan\beta& 0\end{pmatrix},\;\text{and}\;
Y_{E}^{G^{\pm}} = -\sin\beta\begin{pmatrix} 0 & 0 & \kappa_{N}\\ 
							  0& 0 & \kappa \\
						0 & -\bar{\lambda}/\tan\beta& 0\end{pmatrix}.
\end{flalign}

\subsection{Goldstone boson equivalence theorem}
The Goldstone boson equivalence theorem (GBET) gives a relation between $S$-matrix elements of massive vector bosons and unphysical goldstone bosons at high energies through the requirement of tree unitarity \cite{Cornwall:1974km, Lee:1977yc, Lee:1977eg,Riesselmann:1995gv}. In the context of spontaneously-broken gauge theories, this requirement results in useful identities for couplings of goldstone bosons in terms of fermion masses, often simplifying calculations. In the present case it is not immediately obvious how the GBET is satisfied. For instance, the Yukawa matrices of the physical Higgs boson, Eq.~\ref{eq:H_Yukawa}, are clearly different than those appearing for neutral Goldstone bosons, Eq.~\ref{eq:G_couplings} (note the opposite sign appearing with $\bar{\lambda}$). Additionally, the presence of vector-like masses further obscures this equivalence. In this section, we explicitly show the equivalence of Goldstone boson couplings to gauge couplings and fermion masses. This serves as a clarification of these issues in the mass-eigenstate basis, as well as a useful check of gauge invariance of the model.

First, consider the coupling between the neutral Goldstone boson and charged leptons $e_{a}$ and $e_{b}$:
\begin{flalign}
\mathcal{L}_{G}=&-\frac{1}{\sqrt{2}}\bar{\hat{e}}_{La}\lambda^{G}_{e_{a}e_{b}}\hat{e}_{Rb}G + h.c. \\
&=-\frac{1}{2\sqrt{2}}\bar{\hat{e}}_{a}\left[\left(\lambda^{G}_{e_{a}e_{b}} + \left(\lambda^{G\dagger}\right)_{e_{a}e_{b}}\right) +  \left(\lambda^{G}_{e_{a}e_{b}} - \left(\lambda^{G\dagger}\right)_{e_{a}e_{b}}\right)\gamma^{5}\right]\hat{e}_{b}G,
\end{flalign}
which, in terms of lagrangian parameters, can be written as
\begin{flalign}\nonumber
\mathcal{L}_{G} =  -i\frac{\cos\beta}{2\sqrt{2}}\bar{\hat{e}}_{a}\Bigg[&\left((U^{e\dagger}_{L}Y_{E}^{A}U^{e}_{R})_{ab} - (U^{e\dagger}_{R}Y_{E}^{AT}U^{e}_{L})_{ab}\right)\\
&+ \left((U^{e\dagger}_{L}Y_{E}^{A}U^{e}_{R})_{ab} + (U^{e\dagger}_{R}Y_{E}^{AT}U^{e}_{L})_{ab}\right)\gamma^{5}\Bigg]\hat{e}_{b}G.
\end{flalign}
We introduce the following matrices
\begin{flalign}
L=\begin{pmatrix}
	1 & 0 & 0\\
	0 & 1 & 0\\
	0 & 0 & -1
	\end{pmatrix}, \hspace{0.5cm} R = \begin{pmatrix}
	1 & 0 & 0\\
	0 & -1 & 0\\
	0 & 0 & 1
	\end{pmatrix},
\end{flalign}
and note that $Y_{E}^{A}=L\cdot Y_{E} = Y_{E}\cdot R$. Inserting this relation into the vertex factor and applying the unitary relations of $U_{L}^{e}$ and $U_{R}^{e}$ results in
\begin{flalign}
\mathcal{L}_{G} =& \frac{i}{2m_{Z}}\bar{\hat{e}}_{a}\Bigg[\left(m_{e_{b}}- m_{e_{a}}\right)\left[g_{L}^{Zab} + g_{R}^{Zab}\right] + \left(m_{e_{b}} + m_{e_{a}}\right)\left[g_{L}^{Zab} - g_{R}^{Zab}\right]\gamma^{5}\Bigg]\hat{e}_{b}G,
\end{flalign}
where we have identified
\begin{flalign}
\nonumber
g_{L}^{Ze_{a}e_{b}} \equiv &\frac{g}{\cos\theta_W}\left[\left(-\frac{1}{2}+\sin^2\theta_{W}\right)\delta_{ab} + \frac{1}{2}(U_{L}^{e\dagger})_{a5}(U_{L}^{e})_{5b}\right],\\
= &\frac{g}{2c_{W}}\left[\left(-\frac{1}{2} + 2\sin^2\theta_{W}\right)\delta_{ab} - \frac{1}{2}(U_{L}^{e\dagger}LU_{L}^{e})_{ab}\right],
\end{flalign}
and
\begin{flalign}
\nonumber
g_{R}^{Ze_{a}e_{b}}  \equiv& \frac{g}{c_{W}}\left[\sin^2\theta_{W}\delta_{ab} - \frac{1}{2}(U_{R}^{e\dagger})_{a4}(U_{R}^{e})_{4b}\right], \\
= &\frac{g}{2c_{W}}\left[\left(-\frac{1}{2} + 2\sin^2\theta_{W}\right)\delta_{ab} + \frac{1}{2}(U_{R}^{e\dagger}RU_{R}^{e})_{ab}\right].
\end{flalign}

Similar calculations lead to the following relation between the charged Goldstone and $W$ boson couplings
\begin{flalign}
\mathcal{L}_{G^{-}} =&-\frac{1}{2}\bar{\hat{e}}_{a}\Bigg[\left(\lambda^{G^{\pm}}_{e_{a}\nu_{b}} + (\lambda^{G^{\pm}}_{\nu_{b}e_{a}})^{\dagger}\right) + \left(\lambda^{G^{\pm}}_{e_{a}v_{b}} - (\lambda^{G^{\pm}}_{\nu_{b}e_{a}})^{\dagger}\right)\gamma^{5}\Bigg]\hat{\nu}_{b}G^{-}\\
=&\frac{1}{2m_{W}}\bar{\hat{e}}_{a}\Bigg[(m_{\nu_{b}}-m_{e_{a}})\left[g_{L}^{We_{a}\nu_{b}}+g_{R}^{We_{a}\nu_{b}}\right] + (m_{\nu_{b}}+m_{e_{a}})\left[g_{L}^{We_{a}\nu_{b}}-g_{R}^{We_{a}\nu_{b}}\right]\gamma^{5}\Bigg]\hat{\nu}_{b}G^{-},
\end{flalign}
where we have identified
\begin{flalign}
g_{L}^{We\nu}=&\frac{g}{2\sqrt{2}}(U_{L}^{e\dagger})(1+L)U_{L}^{\nu},\\
g_{R}^{We\nu}=&\frac{g}{2\sqrt{2}}(U_{R}^{e\dagger})(1-R)U_{R}^{\nu}.
\end{flalign}

\subsection{Approximate couplings}
\label{approx_formulas}
In this Appendix, we list various approximate formulas for couplings which enter the contributions to $(g-2)_{\mu}$ and relevant constraints on the model from mixing of the muon to VL's. Contributions coming from the $SU(2)$ doublet VL are labeled with index $L$, whereas contributions coming from $SU(2)$ charged and neutral VL singlets are labeled with index $E$ or $N$, respectively, regardless of the hierarchy of masses. We assume that all mixing parameters are of similar order. In this case, the mixing matrices in Eqs.~\ref{eq:mixing_matrices_1} and \ref{eq:mixing_matrices_2} can be written as an expansion in the dimensionless parameters  
\begin{flalign}
\epsilon_{E} =& \frac{v_{d}}{M_{L,E}}\times(\lambda_{L},\lambda_{E},\lambda,\bar{\lambda})\\
\epsilon_{N} =& \frac{v_{u}}{M_{L,N}}\times(\kappa_{N},\kappa,\bar{\kappa}).
\end{flalign}
Thus, in the limit $\lambda_{E}v_{d}, \lambda_{L}v_{d}, \lambda v_{d}, \bar{\lambda} v_{d} \ll M_{L}, M_{E}$, and $\kappa_{N}v_{u}, \kappa v_{u}, \bar{\kappa} v_{u} \ll M_{L}, M_{N}$  the mixing matrices up to order $\mathcal{O}(\epsilon_{E}^{2})$ and $\mathcal{O}(\epsilon_{N}^{2})$ are given by
\begin{equation}
U_{L}^{e}=\begin{pmatrix}
			1-v_{d}^2\frac{\lambda_{E}^2}{2M_{E}^2}& -v_{d}^2\left(\frac{\lambda_{E}}{M_{L}}\frac{\bar{\lambda}M_{E}+\lambda M_{L}}{M_{E}^2-M_{L}^2} - \frac{y_{\mu}\lambda_{L}}{M_{L}^2}\right)& v_{d}\frac{\lambda_{E}}{M_{E}}\\
			v_{d}^2\frac{\bar{\lambda}\lambda_{E}M_{L}-y_{\mu}\lambda_{L}M_{E}}{M_{L}^2M_{E}} & 1-v_{d}^2\frac{(\lambda M_{E} + \bar{\lambda}M_{L})^2}{2(M_{E}^2 - M_{L}^2)^2} & v_{d}\frac{\bar{\lambda} M_{L} + \lambda M_{E}}{M_{E}^2 - M_{L}^2}\\
			-v_{d}\frac{\lambda_{E}}{M_{E}} & -v_{d}\frac{\bar{\lambda}M_{L} + \lambda M_{E}}{M_{E}^2 - M_{L}^2} & 1-v_{d}^2\frac{\lambda_{E}^2}{2M_{E}^2}-v_{d}^2\frac{(\lambda M_{E}+\bar{\lambda}M_{L})^2}{2(M_{E}^2 - M_{L}^2)^2}
			\end{pmatrix},
\end{equation}
\begin{equation}
U_{R}^{e}=\begin{pmatrix}
			1-v_{d}^2\frac{\lambda_{L}^2}{2M_{L}^2}& v_{d}\frac{\lambda_{L}}{M_{L}} &  v_{d}^2\left(\frac{\lambda_{L}}{M_{E}}\frac{\bar{\lambda}M_{L}+\lambda M_{E}}{M_{E}^2-M_{L}^2} + \frac{y_{\mu}\lambda_{E}}{M_{E}^2}\right)\\
			- v_{d}\frac{\lambda_{L}}{M_{L}} & 1-v_{d}^2\frac{\lambda_{L}^2}{2M_{L}^2}-v_{d}^2\frac{(\lambda M_{L}+\bar{\lambda}M_{E})^2}{2(M_{E}^2 - M_{L}^2)^2}& v_{d}\frac{\bar{\lambda} M_{E} + \lambda M_{L}}{M_{E}^2 - M_{L}^2}\\
			v_{d}^2\frac{\lambda_{L}\bar{\lambda}M_{E}-y_{\mu}\lambda_{E}M_{L}}{M_{L}M_{E}^2} & -v_{d}\frac{\bar{\lambda}M_{E}+\lambda M_{L}}{M_{E}^2-M_{L}^2} & 1 - v_{d}^2\frac{(\bar{\lambda}M_{E} + \lambda M_{L})^2}{2(M_{E}^2 - M_{L}^2)^2}
			\end{pmatrix},
\end{equation}
\begin{equation}
U_{L}^{\nu}=\begin{pmatrix}
			1-v_{u}^2\frac{\kappa_{N}^2}{2M_{N}^2} & -v_{u}^2\frac{\kappa_{N}}{M_{L}}\frac{\kappa M_{L} + \bar{\kappa}M_{N}}{M_{N}^2 - M_{L}^2} & v_{u}\frac{\kappa_{N}}{M_{N}}\\
			v_{u}^2\frac{\kappa_{N}\bar{\kappa}}{M_{L}M_{N}} & 1 - v_{u}^2\frac{(\bar{\kappa}M_{L} + \kappa M_{N})^2}{2(M_{N}^2 - M_{L}^2)^2} & v_{u}\frac{\bar{\kappa}M_{L} + \kappa M_{N}}{M_{N}^2 - M_{L}^2}\\ -v_{u}\frac{\kappa_{N}}{M_{N}} & -v_{u}\frac{\bar{\kappa}M_{L} + \kappa M_{N}}{M_{N}^2 - M_{L}^2} &  1 - v_{u}^2\frac{\kappa_{N}^{2}}{2M_{N}^2} - v_{u}^2\frac{(\bar{\kappa}M_{L} + \kappa M_{N})^2}{2(M_{N}^2 - M_{L}^2)^2}
			\end{pmatrix},
\end{equation}
and
\begin{equation}
U_{R}^{\nu} = \begin{pmatrix}
			1 & 0 & 0\\
			0 & 1 - v_{u}^2\frac{(\kappa M_{L} + \bar{\kappa} M_{N})^2}{2(M_{N}^2 - M_{L}^2)^2} & v_{u}\frac{\kappa M_{L} + \bar{\kappa} M_{N}}{M_{N}^2 - M_{L}^2}\\
			0 & - v_{u}\frac{\kappa M_{L} + \bar{\kappa}M_{N}}{M_{N}^2 - M_{L}^2} & 1 - v_{u}^2\frac{(\kappa M_{L} + \bar{\kappa} M_{N})^2}{2(M_{N}^2 - M_{L}^2)^{2}}
			\end{pmatrix}.
\label{eq:A64}
\end{equation}
The above formulas are valid assuming that the mass eigenstates $e_{4}$ and $\nu_{4}$ are mostly doublet-like, while $e_{5}$ and $\nu_{5}$ are mostly singlet-like. This is equivalent to $m_{e_{4}}\simeq M_{L}$, $m_{e_{5}}\simeq M_{E}$, $m_{\nu_{4}}\simeq M_{L}$, and $m_{\nu_{5}}\simeq M_{N}$, and $\lambda v_{d}, \bar{\lambda} v_{d}\ll (M_{E}-M_{L})$ and $\kappa v_{u}, \bar{\kappa} v_{u}\ll (M_{N}-M_{L})$. In the opposite hierarchy of doublets and singlets one can find the corresponding diagonalization matrices by switching the second and third columns of each matrix while simultaneously switching the bottom two entries in each case.

For couplings of the $Z$ boson in this approximation we find
\begin{flalign}
g_{L}^{Z\mu\mu} =& \frac{g}{c_{W}}\left(\left(-\frac{1}{2} + s_{W}^{2}\right) + \frac{v_{d}^{2}\lambda_{E}^{2}}{2M_{E}^2}\right),\\
g_{R}^{Z\mu\mu} =&\frac{g}{c_{W}}\left( s_W^2-\frac{v_d^2 \lambda _L^2}{2 M_L^2}\right),
\end{flalign}
and
\begin{flalign}
g_{L}^{Z\mu L} =& \frac{g}{2c_{W}M_{E}}\frac{(\lambda M_{E} + \bar{\lambda}M_{L})v_{d}^{2}\lambda_{E}}{(M_{E}^2 - M_{L}^2)},\\
g_{R}^{Z\mu L}=&\frac{gv_{d}\lambda_{L}}{2c_{W}M_{L}}\left(-\frac{v_d^2 \left(M_{E} \bar{\lambda }+\lambda  M_L\right){}^2}{2 \left(M_{E}^2-M_L^2\right){}^2}-\frac{v_d^2 \lambda _L^2}{2 M_L^2}+1\right)\\
\simeq& \frac{g}{2c_{W}}\frac{v}{M_{L}}\lambda_{L}\cos\beta,\\
g_{L}^{Z\mu E} =& \frac{gv_{d}\lambda_{E}}{2c_{W}M_{E}}\left(\frac{v_d^2 \left(\bar{\lambda } M_L+\lambda  M_{E}\right){}^2}{2 \left(M_{E}^2-M_L^2\right){}^2}+\frac{\lambda _{E}^2 v_d^2}{2 M_{E}^2}-1\right)\\
\simeq&-\frac{g}{2c_{W}}\frac{v}{M_{E}}\lambda_{E}\cos\beta,\\
g_{R}^{Z\mu E} =& \frac{gv_{d}^2\lambda_{L}}{2c_{W}M_{L}}\frac{(\bar{\lambda}M_{E} + \lambda M_{L})}{(M_{E}^2 - M_{L}^2)},\end{flalign}
where in some formulas we indicate leading order terms in $\epsilon_{E,N}$.

For the corresponding couplings of the $W$ boson to charged and neutral leptons we find
\begin{flalign}
g_{L}^{W\nu\mu} =& \frac{g}{\sqrt{2}}\left( \frac{\bar{\kappa } v_d^2 \kappa _{N} v_{u}^{2} \left(\lambda _{E} \bar{\lambda } M_L-M_{E} \lambda _L y_{\mu }\right)}{M_{E} M_L^3 M_{N}}+\left(1-\frac{\lambda _{E}^2 v_d^2}{2 M_{E}^2}\right) \left(1-\frac{\kappa _{N}^2 v_u^2}{2 M_{N}^2}\right)\right)\\
\simeq&\frac{g}{\sqrt{2}}\left(1 - \frac{\lambda_{E}^{2}v_{d}^{2}}{2M_{E}^{2}} - \frac{\kappa_{N}^{2}v_{u}^{2}}{2M_{N}^{2}}\right),\\
g_{R}^{W\nu\mu} =& \;0,
\end{flalign}
and
\begin{flalign}
g_{L}^{WL\mu} =& \frac{g}{\sqrt{2}}\left( \frac{v_d^2 \left(\lambda _{E} \bar{\lambda } M_L-M_{E} \lambda _L y_{\mu }\right) \left(1-\frac{v_u^2 \left(\bar{\kappa } M_L+\kappa  M_{N}\right){}^2}{2 \left(M_{N}^2-M_L^2\right){}^2}\right)}{M_{E} M_L^2}-\frac{\kappa _{N} v_u^2 \left(1-\frac{\lambda _{E}^2 v_d^2}{2 M_{E}^2}\right) \left(\bar{\kappa } M_{N}+\kappa  M_L\right)}{M_L \left(M_{N}^2-M_L^2\right)}\right)\\
\simeq&\frac{g}{\sqrt{2}}\left(-\frac{v_{u}^{2}\kappa_{N}}{M_{L}}\left(\frac{\bar{\kappa} M_{N} + \kappa M_{L}}{M_{N}^{2} - M_{L}^{2}}\right) + v_{d}^{2}\left(\frac{\lambda_{E}\bar{\lambda}}{M_{E}M_{L}} \right)\right),\\
g_{R}^{WL \mu} =& -\frac{g}{\sqrt{2}}\frac{v_{d}\lambda_{L}}{M_{L}}\left(1-\frac{v_u^2 \left(\bar{\kappa } M_{N}+\kappa  M_L\right){}^2}{2 \left(M_{N}^2-M_L^2\right)^2}\right)\\
\simeq&-\frac{g}{\sqrt{2}}\frac{v}{M_{L}}\lambda_{L}\cos\beta,\\
g_{L}^{WN \mu} =& \frac{g}{\sqrt{2}}\left(\frac{v_d^2 v_u \left(\bar{\kappa } M_L+\kappa  M_{N}\right) \left(\lambda _{E} \bar{\lambda } M_L-M_{E} \lambda _L y_{\mu }\right)}{M_{E} M_L^2 \left(M_{N}^2-M_L^2\right)}+\frac{\kappa _{N} v_u \left(1-\frac{\lambda _{E}^2 v_d^2}{2 M_{E}^2}\right)}{M_{N}}\right)\\
\simeq&\frac{g}{\sqrt{2}}\frac{v}{M_{N}}\kappa_{N}\sin\beta,\\
g_{R}^{WN \mu} =&- \frac{g}{\sqrt{2}}\frac{v_{d}v_{u}\lambda_{L}}{M_{L}}\frac{(\kappa M_{L} + \bar{\kappa}M_{N})}{M_{N}^2 - M_{L}^2}.
\end{flalign}

For the light SM-like Higgs boson, $h$, we obtain
\begin{flalign}
\lambda^{h}_{\mu\mu} =& y_{\mu}\cos\beta \left(1 - \frac{3\lambda_{E}^{2}v_{d}^{2}}{2M_{E}^2} - \frac{3\lambda_{L}^{2}v_{d}^{2}}{2M_{L}^{2}}\right) +\frac{3\cos\beta \lambda _{E} \bar{\lambda } v_d^2 \lambda _L}{M_{E} M_{L}},
	\\
	\lambda^{h}_{\mu L} =& - \frac{\cos\beta \lambda _{E} \bar{\lambda } v_d}{M_{E}}-\frac{v_{d}\cos\beta}{M_{E}^{2}-M_{L}^{2}}\left(\lambda_{E}\bar{\lambda}M_{E} + \lambda_{E}\lambda M_{L}\right),\\\nonumber
	\lambda^{h}_{\mu E} =&\lambda_{E}\cos\beta+  \frac{\cos\beta\lambda  \lambda _{E} \bar{\lambda } v_d^2}{M_{E} M_L} - \frac{\cos\beta \lambda _{E} \bar{\lambda }^2 v_d^2}{M_{E}^2-M_L^2} -\frac{\cos\beta \lambda _{E}^3 v_d^2}{2 M_{E}^2}\\
		\simeq&\lambda_{E}\cos\beta,\\\nonumber
	\lambda^{h}_{L\mu} = & \lambda_{L}\cos\beta + \frac{\cos\beta \lambda  \bar{\lambda } v_d^2 \lambda _L}{M_{E} M_L} + \frac{\cos\beta \bar{\lambda }^2 v_d^2 \lambda _L}{M_{E}^2-M_L^2} -\frac{\cos\beta v_d^2 \lambda _L^3}{2 M_L^2}\\
			\simeq&\lambda_{L}\cos\beta,\\
	\lambda^{h}_{E\mu} =&-\frac{\cos\beta \bar{\lambda } v_d \lambda _L}{M_L}+\frac{v_{d}\cos\beta}{M_{E}^{2}-M_{L}^{2}}\left(\lambda_{L}\bar{\lambda}M_{L} + \lambda_{L}\lambda M_{E}\right).
\end{flalign}
The couplings for the CP-even heavy Higgs $H$ can be obtained simply by replacing one factor of $\cos\beta$ by $\sin\beta$ in the couplings for $h$.

 The couplings of the CP-odd heavy Higgs, A, are given by

\begin{flalign}
i\lambda^{A}_{\mu L} =&\frac{\lambda _{E}\sin\beta \bar{\lambda } v_d}{M_{E}} -\frac{v_{d}\sin\beta}{M_{E}^{2}-M_{L}^{2}}\left(\lambda_{E}\bar{\lambda}M_{E} + \lambda_{E}\lambda M_{L}\right),\\\nonumber
i\lambda^{A}_{\mu E} =&\lambda _{E} \sin\beta + \frac{\lambda  \lambda _{E} \sin\beta \bar{\lambda } v_d^2}{M_{E} M_L}+\frac{\lambda _{E} \sin\beta \bar{\lambda }^2 v_d^2}{M_{E}^2-M_L^2}-\frac{\lambda _{E}^3 \sin\beta v_d^2}{2 M_{E}^2}\\
		\simeq&\lambda_{E}\sin\beta,\\\nonumber
i\lambda^{A}_{L\mu} =&\lambda_{L}\sin\beta +\frac{\lambda  \sin\beta \bar{\lambda } v_d^2 \lambda _L}{M_{E} M_L}-\frac{\sin\beta \bar{\lambda }^2 v_d^2 \lambda _L}{M_{E}^2-M_L^2}-\frac{\sin\beta v_d^2 \lambda _L^3}{2 M_L^2}\\
			\simeq&\lambda_{L}\sin\beta,\\
	i\lambda^{A}_{E\mu} =& \frac{\sin\beta \bar{\lambda } v_d \lambda _L}{M_L}+\frac{v_{d}\sin\beta}{M_{E}^{2}-M_{L}^{2}}\left(\lambda_{L}\bar{\lambda}M_{L} + \lambda_{L}\lambda M_{E}\right).
\end{flalign}

Finally, the couplings for the charged Higgs boson are given by
\begin{flalign}
\nonumber
\lambda^{H^{\pm}}_{L \mu} =&-\lambda_{L}\sin\beta- \frac{\cos\beta \bar{\kappa }^2 v_d \lambda _L v_u}{\left(M_{N}^2-M_L^2\right)}-\frac{\lambda  \sin\beta \bar{\lambda } v_d^2 \lambda _L}{M_{E} M_L} +\frac{\sin\beta v_d^2 \lambda _L^3}{2 M_L^2}\\
			\simeq&-\lambda_{L}\sin\beta,\\
\lambda^{H^{\pm}}_{N \mu} =&	\frac{\cos\beta \bar{\kappa } v_d \lambda _L}{ M_L}-\frac{v_{u}\sin\beta}{M_{N}^{2}-M_{L}^{2}}\left(\lambda_{L}\bar{\kappa}M_{L} + \lambda_{L}\kappa M_{N}\right),\\
\lambda^{H^{\pm}}_{\mu L} =& \frac{\sin\beta \lambda _{E}  \bar{\lambda } v_d}{M_{E}}+\frac{v_{u}\cos\beta}{M_{N}^{2}-M_{L}^{2}}\left(\kappa_{N}\bar{\kappa}M_{N} + \kappa_{N}\kappa M_{L}\right),\\
\lambda^{H^{\pm}}_{\mu N} =&-\kappa_{N}\cos\beta-\frac{\cos\beta \kappa  \lambda _{E} \bar{\lambda } v_d^2}{M_{E} M_L} + \frac{\cos\beta \lambda _{E}^2 v_d^2 \kappa _N}{2 M_{E}^2}\\
		\simeq&-\kappa_{N}\cos\beta.
\end{flalign}

\section{Details of the 2HDM-II-S}
\label{app:SUSY_model}

In this appendix, we calculate the contributions to $(g-2)_{\mu}$ from heavy Higgs bosons and VL's in the 2HDM motivated by supersymmetry (we do not calculate contributions from superpartners). In Table~\ref{table:superfields}, we list the $SU(2)_{L}\times U(1)_{Y}$ charges for the relevant superfields in the calculation of $(g-2)_{\mu}$. All fields are defined as left-handed. The most general superpotential of charged and neutral lepton Yukawa couplings and vector-like masses under these assumptions is 

\begin{table}[!htb]
  \centering
  \begin{threeparttable}
    \begin{tabular}{ccccc|cccccc}\label{tab:SM_fields}&\\ \midrule\midrule
    	\makecell{} &  $l$ & $\bar{e}$ & $H_u$ & $H_d$ &  $L$ & $\bar{L}$ & $\bar{E}$ & $E$ & $N$ & $\bar{N}$\\
	   \cmidrule(l r){1-11}
    \makecell{$SU(2)_L$} & $\mathbf{2}$ & $\mathbf{1}$ & $\mathbf{2}$ & $\mathbf{2}$ & $\mathbf{2}$ & $\mathbf{2}$ & $\mathbf{1}$ & $\mathbf{1} $& $\mathbf{1}$ & $\mathbf{1}$ \\
     \cmidrule(l r){1-11}
      \makecell{$U(1)_Y$} & $-\frac{1}{2}$ & $1$ & $\frac{1}{2}$ & $-\frac{1}{2}$  & $-\frac{1}{2}$ & $\frac{1}{2}$ & $1$ & $-1$ & $0$ & $0$\\
      \midrule\midrule
    \end{tabular}
  \end{threeparttable}
   \caption{Left-handed superfields and their corresponding quantum numbers for SM leptons, Higgs doublets, and VL leptons.}
   \label{table:superfields}
  \end{table}

\begin{flalign}
W_{VLL} =&y_{\mu}H_{d}l\bar{e} + \lambda_{E}H_{d}l\bar{E} + \lambda_{L}H_{d}L\bar{e} + \lambda H_{d}L\bar{E} - \bar{\lambda}H_{u}\bar{L}E\\
&+\kappa_{N}H_{u}l\bar{N} + \kappa H_{u}L\bar{N} - \bar{\kappa}H_{d}\bar{L}N\\
&-M_{L}L\bar{L} + M_{E}E\bar{E} - M_{N}N\bar{N},
\end{flalign}
where the doublet components are labeled as
\begin{equation}
l=\begin{pmatrix} \nu_{\mu}\\ \mu_{L} \end{pmatrix},\hspace{0.25cm} L=\begin{pmatrix} L^{0}_{L}\\ L^{-}_{L}\end{pmatrix}, \hspace{0.25cm} \bar{L}=\begin{pmatrix} \bar{L}^{+}\\ \bar{L}^{0}\end{pmatrix}, \hspace{0.25cm} H_{d}=\begin{pmatrix} H_{d}^{0}\\ H_{d}^{-}\end{pmatrix},  \hspace{0.25cm} H_{u}=\begin{pmatrix} H_{u}^{+}\\ H_{u}^{0}\end{pmatrix},
\end{equation}
and $SU(2)$ doublets are contracted using antisymmetric $\epsilon$, e.g. $H_{d}l=H_{d}^{0}l_{2} - H_{d}^{-}l_{1}=\epsilon^{ab}H_{da}l_{b}$ where $\epsilon^{12} = -\epsilon_{12}$ = +1. Note that $\bar{L}$ field is related to $L_{R}$ introduced in Eq. \ref{eq:Lag} via $\bar{L}=i\sigma_{2}L_{R}^{*}$, and similarly $\bar{E}$ is the chiral supermultiplet which contains $E_{R}^{\dagger}$. In addition, note that the Higgs doublets are defined with opposite hypercharges than in the 2HDM-II-$Z_{2}$. They are related by the field redefinitions $H_{d}=i\sigma^{2}\tilde{H}_{d}^{*}$ and $H_{u}=-i\sigma^{2}\tilde{H}_{u}^{*}$, where the tilde fields are the Higgs doublets defined in the 2HDM-II-$Z_{2}$. Signs of couplings have been chosen so that entries in the mass matrices of charged and neutral leptons have the same sign as in the 2HDM-II-$Z_{2}$ case.

The mass eigenstates, couplings of fermions to gauge and Higgs bosons, and contributions to $(g-2)_{\mu}$ can be calculated in a straightforward way following the procedure detailed in the previous appendix. Note that in the conventions used here the Higgs sector (in alignment limit) is decomposed as 
\begin{flalign}
H_{u}^{0}=&v_{u} + \frac{1}{\sqrt{2}}(h\sin\beta - H\cos\beta) + \frac{i}{\sqrt{2}}(G\sin\beta + A\cos\beta),\\
H_{d}^{0}=&v_{d} + \frac{1}{\sqrt{2}}(h\cos\beta + H\sin\beta) - \frac{i}{\sqrt{2}}(G\cos\beta - A\sin\beta),
\end{flalign}
and
\begin{equation}
\begin{pmatrix} H_{u}^{+}\\ H_{d}^{-*} \end{pmatrix}=R_{\beta}\begin{pmatrix} G^{+}\\ H^{+} \end{pmatrix},
\end{equation}
where
\begin{equation}
R_{\beta}=\begin{pmatrix}\sin\beta & \cos\beta \\
					-\cos\beta & \sin\beta 
					\end{pmatrix},
\end{equation}
and we identify $H^{\pm}=(H^{\mp})^{*}$. With these definitions, the differences in the couplings of gauge and Higgs bosons appear only through $\bar{\lambda}$ and $\bar{\kappa}$ terms, and they are summarized in the main text. Contributions to $(g-2)_{\mu}$ can then be found from the general formulas given in Section (\ref{sec:loops}).

The approximate contributions to $(g-2)_{\mu}$ from $Z$, $W$, and $h$ in the limit of heavy comparable lepton masses are given by

\begin{equation}
\Delta a_{\mu}^{Z} \simeq -\frac{m_{\mu}vc^{3}_{\beta}}{32\pi^{2}}\frac{\lambda_{L}\lambda_{E}\bar{\lambda}}{M_{L}M_{E}}\tan\beta,
\end{equation}
\begin{equation}
\Delta a_{\mu}^{W} \simeq \frac{m_{\mu}vc^{3}_{\beta}}{16\pi^{2}}\frac{\lambda_{L}\lambda_{E}\bar{\lambda}}{M_{L}M_{E}}\tan\beta,
\end{equation}
\begin{equation}
\Delta a_{\mu}^{h} \simeq -\frac{3m_{\mu}vc^{3}_{\beta}}{32\pi^{2}}\frac{\lambda_{L}\lambda_{E}\bar{\lambda}}{M_{L}M_{E}}\tan\beta.
\end{equation}
Assuming $M_{L,E,N}\lesssim m_{A}$, the contributions from $H$, $A$, and $H^{\pm}$ are given by 
\begin{equation}
\Delta a_{\mu}^{H^{\pm}} \simeq -\frac{m_{\mu}vs^{2}_{\beta}c_{\beta}}{96\pi^{2}}\left[\frac{1}{m_{H^{\pm}}^{2}}\left(\frac{\lambda_{L}\lambda_{E}\bar{\lambda}M_{L}}{M_{E}\tan\beta} + \frac{\lambda_{L}\kappa_{N}\bar{\kappa}M_{N}}{M_{L}\tan\beta} + \lambda_{L}\kappa_{N}\kappa\right) + \frac{\lambda_{L}\lambda_{E}\bar{\lambda}}{M_{L}M_{E}\tan\beta}\right],
\end{equation}
\begin{equation}
\Delta a_{\mu}^{H} \simeq \frac{m_{\mu}vs^{2}_{\beta}c_{\beta}}{192\pi^{2}}\left[\frac{1}{m_{H}^{2}}\left(\frac{\lambda_{L}\lambda_{E}\bar{\lambda}(M_{L}^{2}+M_{E}^{2})}{M_{L}M_{E}\tan\beta} + \lambda_{L}\lambda_{E}\lambda\right) + \frac{3\lambda_{L}\lambda_{E}\bar{\lambda}}{M_{L}M_{E}}\left(\frac{2}{\tan\beta} - \tan\beta\right)\right],
\end{equation}
\begin{equation}
\Delta a_{\mu}^{A} \simeq \frac{m_{\mu}vs^{2}_{\beta}c_{\beta}}{192\pi^{2}}\left[\frac{1}{m_{A}^{2}}\left(\frac{\lambda_{L}\lambda_{E}\bar{\lambda}(M_{L}^{2}+M_{E}^{2})}{M_{L}M_{E}\tan\beta} - \lambda_{L}\lambda_{E}\lambda\right) + \frac{3\lambda_{L}\lambda_{E}\bar{\lambda}}{M_{L}M_{E}}\left(\frac{2}{\tan\beta} + \tan\beta\right)\right].
\end{equation}

Compared to the 2HDM-II-$Z_{2}$ version, the loops involving SM bosons are now $\tan\beta$ enhanced. Heavy Higgs contributions contain both $\tan\beta$ enhanced and suppressed terms. Though in the limit when heavy Higgs masses are equal and comparable to heavy lepton masses, the $\tan^{2}\beta$ enhanced contributions cancel in the total contribution. The approximate formulas highly simplify when $M_{L,E,N} = m_{A}$ and vanishing $\kappa$ couplings. In this case we find
\begin{equation}
\Delta a_{\mu}^{H^{\pm}} \simeq - \frac{m_{\mu}vc^{3}_{\beta}}{48\pi^{2}}\frac{\lambda_{L}\lambda_{E}\bar{\lambda}}{M_{L}M_{E}}\tan\beta,
\end{equation}
\begin{equation}
\Delta a_{\mu}^{H,A} \simeq \frac{m_{\mu}vc^{3}_{\beta}}{24\pi^{2}}\frac{\lambda_{L}\lambda_{E}\bar{\lambda}}{M_{L}M_{E}}\tan\beta,
\end{equation}
ignoring terms which cancel between $\Delta a_{\mu}^{H}$ and $\Delta a_{\mu}^{A}$. Note that comparing contributions from $Z$, $W$, and $h$ to those from heavy Higgses, we find that $\Delta a_{\mu}\simeq0$ in this approximation.

\section{Comments on Barr-Zee contributions}
\label{app:BZ}
Two-loop contributions to $(g-2)_{\mu}$ from Barr-Zee (BZ) diagrams can sometimes be competitive with one-loop predictions due to chiral enhancement in the closed fermion loop~\cite{Barr:1990vd}. In the models we have discussed, the chiral enhancement is generated already at the one-loop level, $\Delta a^{\text{1 loop}}_{\mu}\sim m_{\mu}v/M^{2}$, where $M$ is the scale of new physics. The dominant contribution from BZ-type diagrams is generated through diagrams with a neutral Higgs and photon in the internal legs \cite{Ilisie:2015tra}. General formulae for this diagram are given in \cite{Cheung:2009fc,Ilisie:2015tra}. In our notation for Higgs couplings, this contribution from neutral Higgses is given by
\begin{flalign}
\Delta a^{BZ}_{\mu} = \frac{\alpha}{8\pi^{3}}\sum_{\phi=h,H,A}\sum_{a=4,5}\frac{m_{\mu}}{m_{e_{a}}}\left[-\text{Re}(\lambda_{\mu\mu}^{\phi})\text{Re}(\lambda_{e_{a}e_{a}}^{\phi})f(x^{a}_{\phi}) + \text{Im}(\lambda_{\mu\mu}^{\phi})\text{Im}(\lambda_{e_{a}e_{a}}^{\phi})g(x^{a}_{\phi})\right],
\end{flalign}
where $x^{a}_{\phi}=m_{e_{a}}^{2}/m_{\phi}^{2}$ and 
\begin{flalign}
f(\tau)=&\frac{\tau}{2}\int_{0}^{1}dx\frac{1-2x(1-x)}{x(1-x)-\tau}\ln\left(\frac{x(1-x)}{\tau}\right),\\
g(\tau)=&\frac{\tau}{2}\int_{0}^{1}dx\frac{1}{x(1-x)-\tau}\ln\left(\frac{x(1-x)}{\tau}\right).
\end{flalign}

The relative size of the BZ contribution from CP-even Higgses (noting that we work in a CP-conserving 2HDM) compared to the one-loop contribution, Eq. (\ref{eq:Higgs_loops}), is estimated by
\begin{flalign}
\frac{\Delta a^{\phi,BZ}_{\mu}}{\Delta a^{\phi}_{\mu}}\simeq \frac{4\alpha}{\pi}\frac{m_{\phi}^{2}}{m_{e_{a}}^{2}}\frac{-\text{Re}(\lambda_{\mu\mu}^{\phi})\text{Re}(\lambda_{e_{a}e_{a}}^{\phi})}{\text{Re} \left[ \lambda_{\mu e_a}^{\phi} \lambda_{e_a \mu}^{\phi} \right]}\frac{f(x^{a}_{\phi})}{G_{\phi}(x^{a}_{\phi})},
\end{flalign}
where $\phi = h, H$. In the limit of heavy lepton masses, we have \cite{Altmannshofer:2013zba}
\begin{flalign}
f(x)&\overset{x\rightarrow\infty}{\longrightarrow} \frac{1}{3}\ln(x),\\
g(x)&\overset{x\rightarrow\infty}{\longrightarrow} \frac{1}{2}\ln(x),
\end{flalign}
while for the one-loop function
\begin{flalign}
xG_{\phi}(x)&\overset{x\rightarrow\infty}{\longrightarrow} 1.
\end{flalign}
Thus, we can estimate the relative contribution from the BZ diagram with $h$ compared to the corresponding one-loop contribution by
\begin{flalign}
\frac{\Delta a^{h,BZ}_{\mu}}{\Delta a^{h}_{\mu}}\simeq \frac{4\alpha}{3\pi}\frac{m_{\mu}}{v\cos\beta}\ln(x^{a}_{h})\simeq\mathcal{O}(10^{-4} - 10^{-5}),
\label{eq:BZ_higgs}
\end{flalign}
considering $\mathcal{O}(1)$ couplings of VL to the SM Higgs, VL masses up to $20$ TeV, and $\tan\beta=1-50$. We have ignored the overall sign since both signs are possible depending on the signs of $\lambda$'s.

For comparable masses of heavy leptons and the heavy CP-even Higgs, $x^{a}_{H}\simeq 1$, we have $f(x^{a}_{H})\simeq\mathcal{O}(1)$ and $x^{a}_{H}G_{H}\simeq\mathcal{O}(1)$, and the relative contribution is approximated by
\begin{flalign}
\frac{\Delta a^{H,BZ}_{\mu}}{\Delta a^{H}_{\mu}}\simeq \frac{4\alpha}{\pi}\frac{m_{\mu}}{v\cos\beta}\simeq\mathcal{O}(10^{-4} - 10^{-5}),
\end{flalign}
considering $\mathcal{O}(1)$ couplings of VL to $H$ and $\tan\beta=1-50$. The diagram with the CP-odd Higgs gives similar result. In either case, the relative contribution can be roughly another order of magnitude smaller when $M_{L,E}\ll m_{H,A}.$


%\bibliography{ref}

%\bibliographystyle{JHEP}   

\end{document}